\documentstyle[aps]{revtex}
\begin{document}
\title{Disordered Flat Phase and Phase Diagram for Restricted
Solid on Solid Models of Fcc(110) Surfaces}
\vspace{10mm}
\author{Giuseppe Santoro$^{1}$, Michele Vendruscolo$^{1}$,
Santi Prestipino$^{1}$,\\ and Erio Tosatti$^{1,2}$}
\address{
$^1$ Istituto Nazionale per la Fisica della Materia (I.N.F.M.) and
International School for Advanced Studies (S.I.S.S.A.),\\
Via Beirut 4, 34014 Trieste, Italy\\
$^2$ International Centre for Theoretical Physics (I.C.T.P.), Trieste, Italy\\
}
\maketitle
\vspace{5mm}
\begin{abstract}
We discuss the results of a study of restricted solid-on-solid model
hamiltonians for fcc (110) surfaces.
These models are simple modifications of the exactly solvable BCSOS model,
and are able to describe a $(2\times 1)$ missing-row reconstructed surface
as well as an unreconstructed surface.
They are studied in two different ways. The first is
by mapping the problem onto a quantum spin-1/2 one-dimensional
hamiltonian of the Heisenberg type, with competing $S^z_iS^z_j$ couplings.
The second is by standard two-dimensional Monte Carlo simulations.
We find phase diagrams with the following features, which we believe to
be quite generic:
{\rm (i)} two flat, ordered phases (unreconstructed and missing-row
reconstructed); a rough, disordered phase; an intermediate
disordered flat phase, characterized by monoatomic steps, whose physics is
shown to be akin to that of a dimer spin state.
{\rm (ii)} a transition line from the $(2\times 1)$ reconstructed phase to the
disordered flat phase
showing exponents which appear to be close, within our numerical accuracy,
to the 2D-Ising universality class.
{\rm (iii)} a critical (preroughening) line with variable exponents, separating
the unreconstructed phase from the disordered flat phase.
Possible signatures and order parameters of the disordered flat
phase are investigated. \vspace{5mm}
\end{abstract}

%%%%%%%%%%%%%%%%%%%%%%%%%%%%%%%%%%%%%%%%%%%%%%%%%%%%%%%%%%%%%%%%%%%%%%%%%%
\section{Introduction}

Surfaces of fcc metals, in particular (110) faces, display a variety of phase
transitions which have been the subject of considerable experimental and
theoretical work.
Experimentally, the (110) faces of some fcc metals -- such as Au or Pt --
reconstruct at low temperature into a $(2\times 1)$ missing-row (MR) or
related structures, whereas other metals -- like Ag, Ni, Cu, Rh, and Pd --
retain (at least when clean) their bulk--like periodicity.
As temperature is raised, reconstructed surfaces tend to show two separate
transitions: a critical deconstruction transition, and, at a higher
temperature, a Kosterlitz-Thouless roughening transition.\cite{Au_exp,Pt_exp}
On the other hand, unreconstructed surfaces have not been shown, so far, to
reveal a similar two-transition scenario. Only a roughening transition is
well documented in this case.\cite{Cu_exp,Ni_exp}

Based on theoretical considerations and on simulation work, an interesting
and nontrivial interplay has been anticipated between in-plane disordering,
related to deconstruction, and vertical disordering, related to
roughening,\cite{Erio_87} and many other studies have been devoted to the
problem. \cite{Vil_Vil_88,KJT,Vil_Vil_91,MdN_92,Bal_Kar,Mazzeo,Bernasconi}
The situation is, in principle, somewhat different for the two types of
situations, i.e., the unreconstructed and the MR reconstructed cases.
On an fcc (110) surface one can identify two interpenetrating rectangular
sublattices, with origin, say, at $0$ (the ``white'' sublattice) and
A
at $(\sqrt{2}\hat{\bf x}+\hat{\bf y}+\hat{\bf z})a/2$
(the ``black'' sublattice) where $a$ is the lattice parameter,
$\hat{\bf x}=(001)$, $\hat{\bf y}=(1\overline{1}0)$, and $\hat{\bf z}=(110)$.
The unreconstructed (ideal) surface has therefore {\em two\/} $T=0$ ground
states, differing for the sublattice which occupies the top layer
(see Fig.\ \ref{gs:fig}).
den Nijs has argued that, in such a case, the phase diagram should be
qualitatively the same as that of a simple cubic (100) surface.\cite{MdN_90}
In particular, den Nijs,\cite{MdN_90} Jug and Tosatti,\cite{Jug}
Kohanoff {\em et al.},\cite{KJT} and Mazzeo {\em et al.}\cite{Mazzeo}
argued that (110) surfaces like those of Ag and Pd (which do not reconstruct)
are good candidates for realizations of {\em preroughening\/}, a
critical (non-universal) transition from a low temperature ordered phase to
an intermediate {\em disordered flat\/} phase, previously
identified in the context of restricted solid-on-solid models for
simple cubic (100) surfaces.\cite{MdN_Rom}
In terms of the two ground states of the unreconstructed surface, the
preroughening transition can be viewed as a disordering of the surface
due to the proliferation of monoatomic steps (see Fig.\ \ref{defects:fig})
separating terraces with one type of ground state from others where the
other ground state is locally present.
These steps retain, however, an up-down long range order -- stabilized
by a combination of up-up (down-down) step repulsion and entropy -- which
prevents the surface from being rough.\cite{MdN_Rom}

On the $(2\times 1)$ MR reconstructed surfaces, with half of the
$(1{\overline 1}0)$ close-packed rows missing,
the periodicity in the $(001)$ direction is doubled.
The surface has therefore {\em four\/} degenerate $T=0$ ground states,
which can be classified by a clock variable $\theta=0,\pi/2,\pi,3\pi/2$,
according to the ``color'' and the position of the MR in the doubled
unit cell, i.e., determined by which of the four sublattices sits in the top
layer (see Fig.\ \ref{gs:fig}).
The elementary extended defects which one can consider here were
discussed by Vilfan and Villain \cite{Vil_Vil_91} and den Nijs \cite{MdN_92}
(see Fig.\ \ref{defects:fig}).
These are a) {\em steps\/}, which simultaneously change the average height
by $\Delta h=\pm 1$, and the reconstruction variable $\theta$ by
$\Delta\theta =\pi/2$ [{\em clockwise\/} or $(3\times 1)$ steps]
or $\Delta\theta =-\pi/2$ [{\em anticlockwise\/} or $(1\times 1)$ steps],
b) {\em Ising wall defects\/} with $\Delta h=0$ and $\Delta\theta =\pi$,
which can be seen as a tightly bound state between two steps of opposite
sign (up and down), but {\em same} $\Delta\theta$.\cite{MdN_92}

In this framework, den Nijs introduced a phenomenological four state
clock-step model to describe the interplay between reconstruction and
roughening degrees of freedom.\cite{MdN_92}
The model is formulated on a length scale larger than microscopic, through
the introduction of a coarse grained lattice of cells in which
a integer variable $h_{\bf r}$, representing the average height in the cell,
and a clock reconstruction variable $\theta_{\bf r}$ are defined.
A bond in the lattice can be either empty (no defect), or occupied by an up or
down step of either kind, or doubly occupied by an up and down step of the
same kind (equivalent to an Ising wall).
den Nijs found that when $(1\times 1)$ and $(3\times 1)$ steps have
the same energy -- the so-called {\em zero chirality limit\/} -- the
model displays two possible scenarios:
{\em (i)\/} If the energy of an Ising wall $E_w$ is less than roughly twice
the energy of a step $E_s$, temperature drives the system from the ordered
phase to a disordered flat phase through an Ising transition, and then
to a rough phase through a Kosterlitz-Thouless (KT) transition;
{\em (ii)\/} When steps are energetically favored, $E_w>2E_s$,
the system undergoes a single roughening--plus--deconstruction transition,
which is Ising-like for the reconstruction degrees of freedom and
KT-like for the height degrees of freedom.
The disordered flat phase present for $E_w<2E_s$ is quite clearly
characterized by the proliferation of Ising wall defects (their free energy
per unit length goes to zero at the deconstruction). Accordingly, the surface
shows a prevalence of $\theta=0$ and $\theta=\pi$ terraces, say, over
$\theta=\pi/2$ and $\theta=3\pi/2$ ones.
Using the terminology introduced in Ref.\ \cite{Bernasconi}, such a phase
could be called disordered even flat (DEF). It has an obvious non-zero order
parameter which counts the difference in the abundance of $\theta=0,\pi$
terraces over that of $\theta=\pi/2,3\pi/2$ ones, and vanishing only in
the rough phase.
By contrast, when single steps dominate -- i.e., $2E_s<E_w$ -- there is
apparently no mechanism, in this simple model, which may stabilize
the up-down long range order for steps, typical of disordered flat phases.
It has been argued that suitable interactions penalizing the crossing of two
up-up or two down-down steps -- not considered by den Nijs -- could stabilize
such a hypothetical step-dominated disordered flat phase.\cite{Bernasconi}
A disordered flat phase of this kind -- termed DOF in
Ref.\ \cite{Bernasconi} -- should be characterized by an equal abundance of all
types of cells $\theta=0,\pi/2,\pi,3\pi/2$, i.e., by {\em a vanishing of the
order parameter characterizing the DEF phase}.\cite{Bernasconi}

Interestingly, the situation does not change much in the
so-called {\em strong chirality limit\/} considered by den Nijs, i.e.,
when anti-clockwise steps, say, are very costly and thus completely negligible.
In such a case the problem may be mapped onto a one-dimensional fermionic model
containing a Hubbard type on-site step-step interaction $U$ such that
the energy of a Ising wall configuration (doubly occupied site) is
$E_w=2E_s+U$.\cite{Bal_Kar:nota}
For $U<0$, the result is the same as in the zero chirality limit case, i.e.,
a DEF phase is obtained.
When steps dominate -- i.e., for $U>0$ -- two distinct rough phases appear,
the deconstruction transition is no longer of the Ising type, but, again, no
disordered flat phase exists.

The variety of possibilities introduced by the phenomenological models
is thus very rich.
Do {\em microscopic\/} models display just the same, or any new features, one
might ask?
In the light of the previous discussion, the question naturally arises as to
what kind of disordered flat phase (or phases) is realized in simple
solid-on-solid (SOS) models.
The question has been considered by Mazzeo {\em et al.}, who have introduced,
and studied by Monte Carlo simulation, a restricted SOS model which is able
to deal with both unreconstructed and reconstructed situations.\cite{Mazzeo}
Their model -- which we will refer to as ``$K_3$--model'' -- is a simple
modification of the exactly solvable body-centered solid-on-solid model
(BCSOS) \cite{BCSOS:exact} obtained by adding a further neighbor interaction,
which can stabilize the $(2\times 1)$ MR reconstruction if required.
For a reconstructed case, they find two-transitions: a 2D-Ising
deconstruction to a disordered flat phase, and a KT roughening at
a slightly higher temperature.
The unreconstructed case studied also shows two transitions, with a
non-universal critical transition to a disordered flat phase followed by
a KT roughening.\cite{Mazzeo}
The nature of the disordered flat phase present in the model was, however,
not fully characterized.

In the spirit of investigating simple but fully microscopic models,
Santoro and Fabrizio have studied a slightly different modification of the
BCSOS model, which will be referred to as ``$K_4$--model''.\cite{pepin}
They found that this model could be mapped onto a quantum spin-1/2
Heisenberg chain with further-neighbor interactions.\cite{MdN_Rom:ld}
The phase diagram they obtained has two low-temperature ordered phases --
unreconstructed or $(2\times 1)$ MR reconstructed, depending on the
parameters of the interactions -- a high-temperature rough phase, and an
intermediate disordered flat phase.
The physics of the disordered flat phase was argued, by analytical arguments,
to be akin to that of the {\em dimer quantum spin phase\/} studied by
Haldane, \cite{Hal_82}
i.e., a doubly degenerate state which breaks translational invariance
and in which dimer-dimer correlation functions acquire long range order.

In this paper we extend and apply the approach of Ref.\ \cite{pepin}
in such a way as to provide a unified picture of the phase diagram of all
the simple BCSOS-like microscopic models of fcc (110) surfaces studied so far.
First we show that a quantum spin-1/2 hamiltonian also underlies the more
general restricted SOS model where both the couplings considered in Refs.\
\cite{Mazzeo} and \cite{pepin} are included.
The spin-1/2 model is, in all cases, a Heisenberg chain with
$S^z_i S^z_j$ competing antiferromagnetic interactions ranging up to third
neighbors.
Secondly, we unambiguously show that the dimer-phase scenario is realized in
the disordered flat phase, and discuss in detail the surface physics
implications of such a scenario.
Long-range dimer order suggests (see section \ref{dimer:sec}) a
particular type of long-range order for the correlation function between
$(2\times 1)$--steps, and also between {\em local surface maxima\/}.
In particular, one is led to study an order parameter $P_{BW}$ -- previously
introduced by Bernasconi and Tosatti \cite{Bernasconi} -- which
measures the difference in the number of local surface maxima belonging to
the white and the black sublattice of a fcc(110).
Due to ``shadowing'', or to surface lattice contraction, this order
parameter is related to antiphase scattering intensity of He-atoms of
X-rays, respectively, and is thus a quantity of direct interest.
(See section \ref{mc:sec}.)
$P_{BW}$ is studied by finite-size analysis of classical Monte Carlo data, and
found to be non-zero in the disordered flat phase of both
the $K_3$-- and $K_4$--model.
It has a non-monotonic temperature behavior, vanishing only at preroughening
and in the rough phase.

The present paper is organized as follows.
Section \ref{model:sec} introduces the BCSOS-type of models which
we consider.
In Section \ref{mapping:sec} we show in detail how these models may be
mapped onto quantum spin-1/2 chain problems.
In Sections \ref{phase_dia:sec} and \ref{dimer:sec} we discuss the
phase diagrams as well as the physics of a ``dimer'' disordered flat phase.
In Section \ref{mc:sec} we present the results of our Monte Carlo
simulations and discuss possible experimental signatures of a dimer phase.
Section \ref{conclusions:sec}, finally, contains some concluding remarks
as well as a discussion of open problems.

\section{Restricted solid-on-solid models for a fcc (110) metal surface}
\label{model:sec}

The (110) surface of a fcc lattice is comprised of two interpenetrating
rectangular sublattices of lattice constants $a_x=\sqrt{2}a_y$,
which we will conventionally refer to as the white (W) and the black (B)
sublattice.
The surface lattice basis vectors are ${\bf x}=a_x\hat{{\bf x}}$ and
${\bf y}=a_y \hat{{\bf y}}$, where $\hat{{\bf x}}=(001)$, and
$\hat{{\bf y}}=(1{\overline 1}0)$.
In the ideal unreconstructed (110) surface, one of the two sublattices
lies above the other at a distance $a_z=a_y/2$.
Within a solid-on-solid (SOS) framework,\cite{Weeks} one associates to
each site ${\bf r}$ of the lattice a height variable $h_{\bf r}$ which can take
only {\em integer\/} values (take $a_z=1$).
The models we are going to study have an additional restriction, in that the
height difference between each site and its four nearest neighbors
(belonging to the other sublattice) are forced to be $\Delta h=\pm 1$.
A height difference of $0$ is therefore excluded, as well as
larger values of $\Delta h$ (they are energetically more costly).
As a consequence, the values of $h_{\bf r}$ are forced to have opposite parity
on the two sublattices, say even on W and odd on B.
This restriction is typical of the BCSOS model.\cite{BCSOS:exact}
It is probably justified for a metal, where strong inward relaxation
makes the first--second--layer bonds extra strong. On the contrary,
it should not be expected to hold for, say, a rare gas solid (110) face,
where these bonds are in fact weaker.

Our hamiltonian is written as
\begin{equation} \label{model:eqn}
{\cal H}={\cal H}_{\rm BCSOS} + \Delta {\cal H} \;,
\end{equation}
where ${\cal H}_{\rm BCSOS}$ describes interactions between sublattice nearest
neighbors, and $\Delta {\cal H}$ takes into account further-neighbor
interactions which will favor or disfavor reconstructed phases.
Specifically, ${\cal H}_{\rm BCSOS}$ is given by
\begin{equation} \label{mod0:eqn}
{\cal H}_{\rm BCSOS} = K_{2y} \sum_{\bf r} (h_{{\bf r}+{\bf y}}-h_{\bf r})^2
                 \,+\, K_{2x} \sum_{\bf r} (h_{{\bf r}+{\bf x}}-h_{\bf r})^2\;,
\end{equation}
with different coupling strengths in the two directions to take into account
the anisotropy of the surface.
$K_{2y}$ will be always assumed to be {\em positive\/} and is generally the
largest energy in the problem. The correspondent physics is that it is very
costly to create a kink on a tightly packed $(1{\overline 1}0)$ row.
The absolute value of $K_{2x}$, i.e., of the coupling between rows, is instead
much smaller, since atoms in neighbouring rows are only second bulk
neighbours.
For $K_{2x}>0$, the $(110)$ surface is stable in its $(1\times 1)$
unreconstructed form.
If $\Delta {\cal H}=0$ we recover the BCSOS model which is exactly solved
through a mapping to the six vertex model,\cite{BCSOS:exact} and shows
a {\em single transition\/}. This is of the Kosterlitz-Thouless (KT) type,
between a low-temperature ordered (unreconstructed) flat phase and a
high-temperature disordered rough phase.
For $K_{2x}<0$ the surface becomes unstable against $(1{\overline 1}0)$ step
formation.
In this case the final stable state is determined by more distant
interactions, contained in $\Delta{\cal H}$.
As for $\Delta{\cal H}$, two possible simple choices have been made in
the literature, corresponding to what we will refer to as the ``$K_3$--model''
and the ``$K_4$--model''.
The $K_3$--model has been introduced by Mazzeo {\em et al.\/},\cite{Mazzeo}
and is defined by
\begin{equation} \label{modk3:eqn}
\Delta {\cal H}_{(K_3)} = \frac{K_{3}}{2} \sum_{\bf r}
 \left[ (h_{{\bf r}+{\bf x}+{\bf b}} - h_{\bf r})^2 \,+\,
        (h_{{\bf r}+{\bf x}-{\bf y}+{\bf b}} - h_{\bf r})^2 \right] \;,
\end{equation}
with $K_3\ge 0$, and ${\bf b}=({\bf x}+{\bf y})/2$
(see Fig.\ \ref{lattice:fig}).
The introduction of this further-neighbor interaction stabilizes the
$(2\times 1)$ MR reconstructed phase.\cite{Mazzeo}
In fact, it is very easy to check that $K_{2x}<0$, $K_3>0$ stabilizes an
ordered succession of up and down $(1{\overline 1}0)$ steps, which is
precisely the $(2\times 1)$ MR state.
An alternative way of stabilizing the same $(2\times 1)$ MR state
against $(111)$ faceting is realized with the $K_4$--model, whose
$\Delta{\cal H}$ reads
\begin{equation} \label{modk4:eqn}
\Delta {\cal H}_{(K_4)}
= K_{4} \sum_{\bf r} (h_{{\bf r}+2{\bf x}}-h_{\bf r})^2 \;,
\end{equation}
with $K_4\ge 0$. The fourth neighbor interaction in the x-direction has the
effect, once again, of increasing the energy of configurations with
$|h_{{\bf r}+2{\bf x}}-h_{\bf r}|=4$.
This model was originally proposed by Kohanoff and Tosatti,\cite{Kohanoff} and
has been recently studied in detail in Ref.\ \cite{pepin}.
More generally, we could include both types of couplings by taking
\begin{equation} \label{modk3k4:eqn}
\Delta {\cal H} = \Delta {\cal H}_{(K_3)} +  \Delta {\cal H}_{(K_4)} \;.
\end{equation}

In subsequent calculations and simulations,
the lattice will be taken to have $N_c=N_x\times N_y$ primitive cells, i.e.,
$2 N_x \times N_y$ sites.
Periodic boundary conditions are assumed in both directions.
A schematic representation of the lattice and of the interactions considered
is given in Fig.\ \ref{lattice:fig}.

The classical $T=0$ ground states for both models are easy to work out
as a function of the dimensionless ratio ${\cal K}=K_{2x}/K_3$ or
${\cal K}=K_{2x}/K_4$.
For both models one finds that ${\cal K}>0$ corresponds to an unreconstructed
ground state [or $(1\times 1)$], whereas for $-4<{\cal K}<0$ the ground state
is $(2\times 1)$ MR reconstructed.\cite{Mazzeo}
For the sake of completeness, we mention that, for the $K_3$--model, the
ground state degenerates into an infinite $(111)$ large facet as soon
as ${\cal K}<-4$.\cite{Mazzeo}
For the $K_4$--model, on the contrary, an infinite $(111)$ facet sets is
only for ${\cal K}<-8$, and there is a whole series of intermediate regions
[$(12-8n)/(n-1)<{\cal K}<(20-8n)/(n-2)$ with $n\ge 3$] where the
ground state is $(n\times 1)$ MR reconstructed.
In the following we will be interested exclusively in the region of parameter
space where the interplay between unreconstructed and $(2\times 1)$
MR reconstructed phases takes place, hence ${\cal K}>-4$.

Neglecting adatoms, vacancies and (as necessary in SOS models) overhangs,
the defects which should play a role in the disordering and roughening
transitions are unbound steps and bound pairs of steps,
i.e., Ising domain walls.
Fig.\ \ref{defects:fig} illustrates the most relevant defects for both a
$(1\times 1)$ and a $(2\times 1)$ surface.
The ground state energies of these defects are given, for both the $K_3$ and
the $K_4$--model, in the table.
It is worth noting that the $K_3$--model has defects whose energy goes to zero
as $K_{2x}\rightarrow 0$.
These are the $(2\times 1)$--step and the $(2\times 1)^{\ast}$--wall in the
$(1\times 1)$ case, and the anticlockwise--step
and the Ising wall in the $(2\times 1)$ reconstructed case.
Since this leads to zero-point entropy, the $K_{2x}\rightarrow 0$ region
is therefore somewhat unphysical for the $K_3$--model, where one might
expect disorder to occur at very low
temperatures.\cite{k3_k3k4_diff:nota}
As a second point, we observe that in the $K_4$--model a combination
of two Ising walls wins against combinations involving clockwise--steps
for ${\cal K}=K_{2x}/K_4>-1$, while it always wins against
anticlockwise--steps.
In principle, therefore, a DEF (wall dominated) phase seems to be
plausible for $-1<{\cal K}<0$ in the $K_4$--model.
Later on we will present results %obtained for ${\cal K}=-0.56$,
which show how ground state defect energy considerations can be somewhat
misleading: the disordered phase obtained does not have the features of an
ideal DEF.

\section{Mapping into a quantum spin-1/2 chain}
\label{mapping:sec}

An elegant and convenient way of studying the temperature phase diagram of the
classical two-dimensional models introduced in the previous section consists
in mapping them into one-dimensional quantum problems.\cite{MdN_Rom:ld}
The general procedure is well known,\cite{Kogut,MdN_Rom:ld,denNijs:rew} but
we review it here for our specific case, and for the reader's convenience.

The method consists in viewing the y-direction of the lattice as the
(imaginary) time direction of an appropriate 1D quantum problem, different
row configurations in the x-direction being viewed as subsequent time-slices
for the quantum problem.
The physical requirement which will turn out to be important is that the
coupling $K_{2y}$ in the y-direction is strong, while the other couplings
are much weaker (strong anisotropy limit).
The starting point for the mapping is a transfer matrix formulation of
the classical partition function ${\cal Z}$.
The notation ${\bf r}=(i,j)$ (and $h_{\bf r}=h^{(j)}_i$) for the lattice sites
used in the present section is illustrated in Fig.\ \ref{lattice:fig}:
the value of $j$, the time-slice index, is the same along each dashed
zig-zag line shown in Fig.\ \ref{lattice:fig};
within each zig-zag line, the W and the B sublattices are characterized,
respectively, by even and odd values of $i$.
The classical partition function is given by
\begin{equation} \label{PART_FUNC:eqn}
{\cal Z} = \sum_{\{h^{(j)}_i\}} e^{-\beta{\cal H}} =
\sum_{h^{(1)}} \cdots \sum_{h^{(N_y)}}
\langle h^{(1)}| \hat{T} |h^{(N_y)}\rangle \cdots
\langle h^{(3)}| \hat{T} |h^{(2)}\rangle
\langle h^{(2)}| \hat{T} |h^{(1)}\rangle \;,
\end{equation}
where $|h^{(j)}\rangle=\{h^{(j)}_i:i=1,\cdots,N_s\}$ is the $j$-th
{\em row configuration\/} (a dashed zig-zag line in Fig.\ \ref{lattice:fig},
containing $N_s=2N_x$ sites), and $\hat{T}$ is the classical transfer matrix.
Periodic boundary conditions have been used in the y-direction, and are
understood in the x-direction.
It is also understood that configurations differing by a uniform shift of
the heights should be included only once in the partition sum.
For the models we are considering, the transfer matrix elements read:
\begin{equation} \label{TMAT_EX:eqn}
\langle h^{(j+1)}| \hat{T} |h^{(j)}\rangle \,=\, B_{nn} \; \exp{ \{
-\beta K_{2y}\sum_{i=1}^{N_s} (h^{(j+1)}_{i}-h^{(j)}_{i})^2 \,
-\beta K_{2x}\sum_{i=1}^{N_s} (h^{(j)}_{i+2}-h^{(j)}_{i})^2 \} } \;,
\end{equation}
where $B_{nn}$ is the Boltzmann weight contribution due to further
neighbor interactions.
For the $K_3$--model, $B_{nn}$ is given by:
\begin{eqnarray} \label{BNN_K3:eqn}
B_{nn}^{(K_3)} &\,=\,& \exp{ \{ -\frac{\beta K_{3}}{2}
\sum_{i\; \mbox{even}}^{N_s} [(h^{(j)}_{i+3}-h^{(j)}_{i})^2
+ (h^{(j)}_{i+3}-h^{(j+1)}_{i})^2] \} } \nonumber \\
&&\; \times \exp{ \{ -\frac{\beta K_{3}}{2}
\sum_{i\; \mbox{odd}}^{N_s} [(h^{(j)}_{i+3}-h^{(j)}_{i})^2
+ (h^{(j+1)}_{i+3}-h^{(j)}_{i})^2] \} } \;,
\end{eqnarray}
whereas for the $K_4$--model the result is:
\begin{equation} \label{BNN_K4:eqn}
B_{nn}^{(K_4)} \,=\,
\exp{ \{ -\beta K_{4}\sum_{i=1}^{N_s} (h^{(j)}_{i+4}-h^{(j)}_{i})^2 \} } \;.
\end{equation}
For the model in which both couplings are present, one clearly has,
\begin{equation} \label{BNN:eqn}
B_{nn} = B_{nn}^{(K_3)} B_{nn}^{(K_4)} \;.
\end{equation}
Notice that in the partition function, Eq.\ \ref{PART_FUNC:eqn},
it is implicitly assumed that the configurations included have
to fulfil the BCSOS constraint $\Delta h=\pm 1$ for nearest neighbors.
As a consequence, within each row we must
have $h^{(j)}_{i+1}-h^{(j)}_{i}=\pm 1$. Therefore, we can associate to any row
configuration $|h^{(j)}\rangle$ a state $|j\rangle$ in the Hilbert space
of a quantum spin-1/2 chain (of length $N_s$) by the relationship
\begin{eqnarray} \label{RELAT_H_SPIN:eqn}
|h^{(j)}\rangle &\longrightarrow &
|j\rangle=|S_1,S_2,\cdots,S_{N_s}\rangle \nonumber \\
S_i &=& \frac{1}{2} (h^{(j)}_{i+1}-h^{(j)}_{i})  \;.
\end{eqnarray}
(In doing so we lose information on the absolute height of the surface,
which is however irrelevant in a static context.)
Fig.\ \ref{gs:fig} illustrates the explicit mapping of the $(1\times 1)$ and
$(2\times 1)$ ground states in terms of spin configurations.

The idea is now to reproduce the Boltzmann factors appearing in the
matrix elements of the classical transfer matrix
$\langle h^{(j+1)}|\hat{T}|h^{(j)}\rangle$ by a suitable quantum
operator $T_Q$ in the spin Hilbert space, i.e.,
\begin{equation} \label{RELAT_T_TQ:eqn}
\langle h^{(j+1)}| \hat{T} |h^{(j)}\rangle =
\langle j+1| T_Q |j\rangle \;,
\end{equation}
where $|j\rangle$ and $|j+1\rangle$ are the quantum spin states corresponding
to
$|h^{(j)}\rangle$ and $|h^{(j+1)}\rangle$, respectively.
In certain cases, the exact expression for the quantum operator
$T_Q$ can be worked out quite easily.
$T_Q$ for the $K_4$--model has been derived in Ref.\ \cite{pepin}.
The exact $T_Q$ is, however, of no practical use, being a
product of non-commuting terms involving spin-1/2 operators.
The crucial step which makes the whole mapping useful
is the so-called {\em time-continuum limit\/} or {\em strong anisotropy
limit\/}.
Physically, one assumes that the ``time'' direction coincides with
the ``hard'' direction of the classical problem, i.e., that the coupling
in the y direction is much stronger than the other couplings.
This is plausibly the case for the fcc(110) surface, where
the $\hat{{\bf y}}=(1{\overline 1}0)$ direction is hard, and the
$\hat{{\bf x}}=(001)$ direction is soft, as discussed above.
Anisotropy, moreover, is not expected to play any role in the
qualitative shape of the phase diagram.\cite{MdN_Rom}

In the strong anisotropy limit the quantum operator $T_Q$ will reduce to the
imaginary-time evolution operator $e^{-\tau H_S}$ for a suitable hamiltonian
$H_S$, with $\tau\rightarrow 0$.
To find $H_S$, assume $\beta K_{2y}$ to be large, so that
$e^{-4\beta K_{2y}}=(J/2)\tau$, $\tau$ being a small quantity (and $J$ of
order one).
Assume also all the other couplings to be small and proportional to $\tau$,
i.e., $\beta K_{2x}\propto\tau$, $\beta K_{3}\propto\tau$, and
$\beta K_{4}\propto\tau$.
We need a spin hamiltonian $H_S$ such that Eq.\ \ref{RELAT_T_TQ:eqn}
is verified with
\begin{equation} \label{TCL:eqn}
T_Q\approx e^{-\tau H_S}\approx 1-\tau H_S + O(\tau^2) \;,
\end{equation}
{\em up to first order in the small quantity\/} $\tau$.
A {\em diagonal\/} matrix elements of $\hat{T}$ reads, using
Eqs.\ \ref{TMAT_EX:eqn} to \ref{BNN:eqn},
\begin{eqnarray} \label{T_DIAG:eqn}
\langle h^{(j)}| \hat{T} |h^{(j)}\rangle
&\,=\,& 1 - \beta K_{2x} \sum_i (h^{(j)}_{i+2}-h^{(j)}_{i})^2
-\beta K_3 \sum_i (h^{(j)}_{i+3}-h^{(j)}_{i})^2 \nonumber \\
&& \hspace{42mm} -\beta K_4 \sum_i (h^{(j)}_{i+4}-h^{(j)}_{i})^2 \,+\,
O(\tau^2)
\end{eqnarray}
where we have expanded all the exponentials up to first order in small
quantities ($\propto\tau$).
The first requirement for $H_S$ is that its diagonal terms must
give the same result, i.e., using Eqs.\ \ref{TCL:eqn}, \ref{T_DIAG:eqn} and
\ref{RELAT_H_SPIN:eqn},
\begin{eqnarray} \label{MAP_0:eqn}
\langle j| T_Q |j\rangle &\,\approx\,& 1 - \tau \langle j | H_S |j\rangle
\,+\, O(\tau^2) \nonumber \\
&\,=\,& 1 - 4\beta K_{2x} \sum_i (S_i+S_{i+1})^2
-4\beta K_3 \sum_i (S_i+S_{i+1}+S_{i+2})^2  \nonumber \\
&& \hspace{4mm}
-4\beta K_4 \sum_i (S_i+S_{i+1}+S_{i+2}+S_{i+3})^2 \,+\, O(\tau^2) \;.
\end{eqnarray}
An {\em off-diagonal\/} matrix element of $\hat{T}$ must contain a Boltzmann
factor $e^{-4\beta K_{2y}}=(J/2)\tau$, for each site $i$ such that
$h^{(j+1)}_i=h^{(j)}_i\pm 2$. Therefore, up to first order in $\tau$ we need
to consider only row configurations $h^{(j+1)}$ which differ from $h^{(j)}$
only at a single site $i$.
Let $\{\cdots,h_{i-1},h_i,h_{i+1},\cdots\}$ be the local configuration
of row $j$ around such a site $i$.
It is easy to realize that, in order to satisfy the BCSOS constraint with
$h^{(j+1)}_i=h^{(j)}_i\pm 2$, and $h^{(j+1)}_k=h^{(j)}_k$ for $k\ne i$, the
only possibility is to have $h^{(j)}_{i-1}=h^{(j)}_{i+1}=h^{(j)}_i\pm 1$,
i.e.,
\[ |h^{(j)}\rangle =
\{\cdots,h_i\pm 1,h_i,h_i\pm 1,\cdots\} \; \longrightarrow \;
|h^{(j+1)}\rangle =
\{\cdots,h_i\pm 1,h_i\pm 2,h_i\pm 1,\cdots\} \;. \]
In terms of the corresponding spin configurations, this simply leads to a
{\em spin flip\/} of the spins at sites $i-1$ and $i$,
\[ |j\rangle =
|\cdots,S_{i-1}=\mp\frac{1}{2},S_i=\pm\frac{1}{2},\cdots \rangle
\; \longrightarrow \; |j+1\rangle =
| \cdots,S_{i-1}=\pm\frac{1}{2},S_i=\mp\frac{1}{2},\cdots \rangle \;. \]
The corresponding $\hat{T}$-matrix element reads, up to first order in $\tau$,
\begin{eqnarray} \label{MAP_1:eqn}
\langle h^{(j+1)}| \hat{T} |h^{(j)}\rangle
&=& e^{-4\beta K_{2y}} \, (1 \,+\, O(\tau) ) \nonumber \\
&\approx& -\tau\langle j+1|H_S|j\rangle \;.
\end{eqnarray}
It is easy to verify, in conclusion, that the correct form
of $H_S$ verifying Eqs.\
\ref{MAP_0:eqn} and \ref{MAP_1:eqn} is given, neglecting an overall
constant, by
\begin{equation} \label{SPIN_MOD:eqn}
H_S = -\frac{J}{2} \sum_{i=1}^{N_s} [S^+_i S^-_{i+1} + S^-_i S^+_{i+1}]
+ \sum_{i=1}^{N_s} [J_z S^z_i S^z_{i+1} + J_2 S^z_i S^z_{i+2} +
J_3 S^z_i S^z_{i+3} ] \;,
\end{equation}
where the spin couplings are related to the original couplings as follows:
\begin{eqnarray} \label{COUP_RELAT:eqn}
\tau J   &=& 2 \exp{(-4\beta K_{2y})} \nonumber \\
\tau J_z &=& 8 \beta (K_{2x} + 2K_3 + 3K_4) \nonumber \\
\tau J_2 &=& 8 \beta (K_3 + 2K_4) \nonumber \\
\tau J_3 &=& 8 \beta K_4 \;.
\end{eqnarray}
Indeed, the spin-flip part of $H_S$ reproduces the off-diagonal matrix
element in Eq.\ \ref{MAP_1:eqn}, whereas the $S^z S^z$ terms
give rise to the correct diagonal matrix element in Eq.\ \ref{MAP_0:eqn}.

It is well known that this kind of mapping is such that the free energy
per site of the classical problem -- given by the maximum eigenvalue of the
transfer matrix -- is related to the ground state energy per site of the
one-dimensional quantum problem, $\beta f=\tau\epsilon_{GS}$.\cite{Kogut}
The temperature clearly enters through the spin couplings, see Eq.\
\ref{COUP_RELAT:eqn}, so that any genuine singularity of the classical
free energy versus temperature can be seen as a ground state energy
singularity for the quantum problem as a function of the couplings
$J_z/J$, $J_2/J$ and $J_3/J$.
Moreover, temperature averages for correlation functions of the classical
problem can be likewise rewritten in the form of ground state averages
for the corresponding quantum correlation function.\cite{Kogut}
In summary, to obtain information about the temperature phase diagram of
the classical model one studies the {\em ground state phase diagram\/}
of the spin chain model. \cite{PHASE_DIAG:nota}

Before entering into the discussion of the phase diagram, let us clarify that
the quantum mapping not only gives the correct {\em critical behavior\/} of the
transitions (if anisotropy is not ``relevant'' in the renormalization
group sense), but provides also {\em quantitative\/} results on the
transition temperatures that are expected to be quite good even if
the anisotropy is in reality only weak.
As a simple check of this point, consider the exactly solvable BCSOS case,
whose transition temperature is given by \cite{BCSOS:exact}
\begin{equation} \label{BCSOS_TCE:eqn}
e^{-4\beta_c K_{2y}} + e^{-4\beta_c K_{2x}} = 1 \;.
\end{equation}
The BCSOS model maps -- see Eq.\ \ref{COUP_RELAT:eqn} -- into the nearest
neighbor XXZ Heisenberg chain, which is known to have a KT transition at
the isotropic point $J_z=J$.\cite{Hal_82}
Using Eq.\ \ref{COUP_RELAT:eqn}, this readily implies a predicted transition
temperature $\beta^{(Q)}_c$ satisfying
\begin{equation} \label{BCSOS_TCQ:eqn}
e^{-4\beta^{(Q)}_c K_{2y}} = 4\beta^{(Q)}_c K_{2x} \;.
\end{equation}
Fig.\ \ref{tc_bcsos:fig} shows both the exact (solid line) and the quantum
mapping transition temperature (dashed line) for the BCSOS as a function of
the anisotropy ratio $K_{2x}/K_{2y}$.
The results agree within a few percents even for rather weak anisotropies,
such as $K_{2x}/K_{2y}=0.2$, and remain reasonable all the way to the
full isotropic case, $K_{2x}/K_{2y}=1$.

\section{Phase diagrams}
\label{phase_dia:sec}

\subsection{Phase diagram of the spin-1/2 chain}
\label{spin_phase_dia:sec}

The spin chain hamiltonian corresponding to our modified BCSOS model
is a Heisenberg XXZ model with a second neighbor and a (less important) third
neighbor $S^z_i S^z_{j}$ coupling
\begin{equation} \label{SPIN_MOD_J1J2J3:eqn}
H_S = -\frac{J}{2} \sum_{i=1}^{N} [S^+_i S^-_{i+1} + S^-_i S^+_{i+1}]
+ \sum_{i=1}^{N}
[J_z S^z_i S^z_{i+1} + J_2 S^z_i S^z_{i+2} + J_3 S^z_i S^z_{i+3}] \;.
\end{equation}
Qualitatively -- and for not too large values of the couplings $J_z,J_2,J_3$ --
the physics of such a model is closely related to that of the Heisenberg chain
with spin-isotropic second neighbor interactions
\begin{equation} \label{HAL_MOD:eqn}
H = \frac{J}{2} \sum_{i} [S^+_i S^-_{i+1} + S^-_i S^+_{i+1}]
+ \sum_{i} [J_z S^z_i S^z_{i+1} + J_2 {\vec S}_i \cdot {\vec S}_{i+2}] \;,
\end{equation}
which we will refer to, in the following, as the
$J-J_2$--model.\cite{kin_sign:nota}
Haldane has discussed the qualitative phase diagram of the $J-J_2$--model
in the context of the Luttinger liquid framework, with special emphasis on
the role played by umklapp processes in the underlying spinless
fermion problem.\cite{Hal_82}
(For a detailed quantitative analysis see Ref.\ \cite{Nomura})
For the spin model in Eq.\ \ref{SPIN_MOD_J1J2J3:eqn}, the discussion goes
along similar lines.
For completeness we will give in subsection \ref{details:sec} some details
of this analysis based on standard techniques of one dimensional
systems.\cite{Hal_82,Solyom}

Even a simple mean field treatment, however, is quite instructive about the
nature of the ordered phases which are to be expected.
The starting point is to perform a Wigner-Jordan transformation from spin
variables to spinless fermion operators $c_i$, i.e.,
$S^z_i=c^{\dagger}_ic_i-1/2$, $S^+_i=c^{\dagger}_i\exp{(i\pi\sum_{j<i}n_j)}$.
Neglecting constants and terms proportional to the total number of fermions,
the spin model is then rewritten as the following {\em spinless fermion\/}
model
\begin{eqnarray} \label{FERMION_H_1:eqn}
H_F &=& -t \sum_{i=1}^{N} [c^{\dagger}_i c_{i+1} + c^{\dagger}_{i+1} c_{i} ]
+ \sum_{i=1}^{N} [J_z n_i n_{i+1} + J_2 n_i n_{i+2} + J_3 n_i n_{i+3}]
\nonumber \\
&=& \sum_{k}^{BZ} \epsilon_k c^{\dagger}_k c_k + \frac{1}{N} \sum_q^{BZ}
v(q) \rho(q) \rho(-q) \;,
\end{eqnarray}
where $c_k=N^{-1/2}\sum_j e^{-ikaj} c_j$, with $k$ belonging to the first
Brillouin zone $[-\pi/a,\pi/a]$ ($BZ$), and $\rho(q)$ is the Fourier transform
of the fermion density operator, $\rho(q)=\sum_k c^{\dagger}_{k}c_{k+q}$.
Here $\epsilon_k=-2t\cos{(ka)}$ with $t=J/2$, and $v(q)$ is the Fourier
transform of the interaction potential,
$v(q)=J_z\cos{(qa)}+J_2\cos{(2qa)}+J_3\cos{(3qa)}$.
Since $\sum_i S^z_i=\sum_i n_i - N/2$, zero total magnetization for the spin
system implies half-filling for the fermions, i.e., a Fermi surface
consisting of two Fermi points at $\pm k_F$, with $k_Fa=\pi/2$.
In absence of interaction ($J_z=J_2=J_3=0$, i.e., the XY spin chain) we have
a simple free-fermion problem.
The two Fermi points induce nesting with a wave-vector $2k_F=\pi/a$, a
hint that the system would tend to open-up a gap at the Fermi surface
by developing long-range order (LRO) with wave-vector $\pi/a$ and making the
average $A_k = \langle c^{\dagger}_{k+\pi/a} c_k \rangle$ different from zero.
A standard mean-field factorization of the quartic term in
Eq.\ \ref{FERMION_H_1:eqn}, assuming
$\langle c^{\dagger}_{k+\pi/a} c_k \rangle \ne 0$, leads to the
following mean-field hamiltonian (we take $a=1$ from now on)
\begin{equation} \label{FERMION_HMF:eqn}
H_F^{\rm MF} = \sum_{k}^{RBZ} \tilde{\epsilon}_k [
c^{\dagger}_{k} c_k - c^{\dagger}_{k+\pi} c_{k+\pi}] +
[\Delta_k c^{\dagger}_{k} c_{k+\pi} + H.c.] \;,
\end{equation}
where $\tilde{\epsilon}_k=\epsilon_k
- (2/N) \sum_{k'\in BZ} v(k-k')\langle c^{\dagger}_{k'} c_{k'}\rangle$,
and $\Delta_k = (2/N) \sum_{k'\in BZ} [v(\pi)-v(k-k')]
\langle c^{\dagger}_{k'+\pi} c_{k'}\rangle$.
Here $RBZ$ stands for the reduced magnetic Brillouin zone $(-\pi/2,\pi/2)$.
Diagonalizing the simple $2\times 2$ problem appearing in $H_F^{\rm MF}$,
one readily finds that the ``anomalous'' average
$A_k=\langle c^{\dagger}_{k+\pi} c_k\rangle$ is simply related to $\Delta_k$,
$A_k=-\Delta_k/(2E_k)$ with $E_k=\sqrt{\tilde{\epsilon}^2_k + |\Delta_k|^2}$,
and that $\tilde{\epsilon}_k$ and $\Delta_k$ have to obey the self-consistency
conditions
\begin{eqnarray} \label{self_consistency:eqn}
\tilde{\epsilon}_k &=& \epsilon_k + \frac{1}{N}\sum_{k'}^{RBZ}
[v(k-k')-v(k-k'+\pi)] \frac{ \tilde{\epsilon}_{k'} }{E_{k'}}
\nonumber \\
\Delta_k &=& -\frac{1}{N} \sum_{k'}^{RBZ} \frac{1}{E_{k'}}
\{ [v(\pi)-v(k-k')] \Delta_{k'} + [v(\pi)-v(k-k'+\pi)] \Delta^*_{k'} \} \;.
\end{eqnarray}
Let us consider, for definiteness, the case $J_3=0$. Solving the
self-consistency equations \ref{self_consistency:eqn}, one finds that if
$v(\pi)=J_2-J_z$ is sufficiently negative ($J_2$ less than $\approx 0.4 J_z$),
$\Delta_k$ is {\em real\/} and has the form
$\Delta_k=\delta_0 + \delta_2 \cos{(2k)}$.
Upon increasing $J_2$, a transition occurs to a phase in which $\Delta_k$
is {\em purely imaginary\/}, $\Delta_k = i\delta_1 \sin{(k)}$.
The transition appears to be first order in mean-field.
To understand the meaning of the two phases, consider the average values
of the fermion density $\langle n_i\rangle$, and of the bond kinetic energy
$\langle c^{\dagger}_{i} c_{i+1} + c^{\dagger}_{i+1} c_{i} \rangle$.
A simple calculation shows that
\begin{eqnarray}
\langle n_i \rangle &=& \frac{1}{2} \,-\, (-1)^i \,\frac{1}{N} \sum_k^{RBZ}
\frac{ {\rm Re} \Delta_k }{E_k} \nonumber \\
\langle c^{\dagger}_{i} c_{i+1} + c^{\dagger}_{i+1} c_{i} \rangle &=&
{\rm const} \,+\, (-1)^i \, \frac{2}{N} \sum_k^{RBZ} \sin{(k)}
\frac{ {\rm Im} \Delta_k }{E_k} \;.
\end{eqnarray}
The phase with $\Delta_k$ real (small $J_2$) is therefore a $2k_F$
{\em site-centered charge density wave\/} (CDW) (i.e., a N\'eel phase with
$\uparrow\downarrow\uparrow\downarrow$ LRO in the spin language, or
an ordered fcc(110) face in the surface language),
whereas the phase at larger $J_2$, with $\Delta_k$ purely imaginary, is
a $2k_F$ {\em bond-centered charge density wave\/}.
We can picture the latter by assuming that on every other bond the fermions
are in a state which maximizes the kinetic energy of the bond,
\begin{equation} \label{Bond_CDW:eqn}
\Psi_{\rm Bond-CDW} = \prod_{i \,{\rm even}}
\frac{ (c^{\dagger}_i + c^{\dagger}_{i+1}) }{\sqrt{2}} |0\rangle \;,
\end{equation}
as opposed to the ideal site-CDW state
(the N\'eel state $\uparrow\downarrow\uparrow\downarrow$)
\begin{equation} \label{Site_CDW:eqn}
\Psi_{\rm Site-CDW} = \prod_{i \,{\rm even}}
c^{\dagger}_i |0\rangle \;.
\end{equation}
The spin interpretation of the bond-CDW state is quite obviously a
dimerized spin state with every other bond engaged in a singlet,
$(\uparrow\downarrow-\downarrow\uparrow)$.\cite{kin_sign:nota}
Unlike the site-CDW, where every second neighbor is occupied and pays an
energy $J_2$, a bond-CDW reduces the second-neighbor average occupancy
to about $1/2$, and is thus favoured upon increasing $J_2$.
As will be discussed in detail, this spin dimer phase corresponds
to a disordered flat phase in the surface language.
Clearly, for very large $J_2$, the system will eventually prefer to minimize
second-neighbor occupancies by forming a site CDW of double periodicity
(i.e., a $k_F$-CDW), which we can picture as
\begin{equation} \label{2_CDW:eqn}
\Psi_{\rm k_F-CDW} = \prod_{i=4n }
c^{\dagger}_i c^{\dagger}_{i+1} |0\rangle \;.
\end{equation}
(Such a state corresponds to $\uparrow\uparrow\downarrow\downarrow$ LRO in spin
language, or a $(2\times 1)$ MR reconstructed face in the surface language.)
This phase can be included in a mean-field treatment by allowing, in the
factorization of the quartic term, anomalous averages
of the type $\langle c^{\dagger}_{k\pm \pi/2} c_k \rangle$, as well as
the previous one $\langle c^{\dagger}_{k+\pi} c_k \rangle$.
%The resulting $4\times 4$ problem can be easily handled numerically.

A finite-size scaling study of the spin model readily confirms most of
the qualitative features of the mean-field phase-diagram.
A quantitative phase diagram for the spin model corresponding
to the $K_3$--model, i.e., Eq.\ \ref{SPIN_MOD_J1J2J3:eqn} with $J_3=0$,
is presented in Fig.\ \ref{phase_j1j2:fig}.
The procedure to obtain such a phase diagram from a finite-size scaling
study of chains up to $N=28$ sites,\cite{Sch_Zim} was
described in detail in Ref.\ \cite{pepin}.
(See also Ref.\ \cite{Nomura}.)
A similar phase diagram for the spin chain corresponding
to the $K_4$--model, i.e.,  Eq.\ \ref{SPIN_MOD_J1J2J3:eqn} with
$J_3=J_2/2$, was presented in Fig.\ 1 of Ref.\ \cite{pepin}.
For the purpose of a general discussion, we reproduce in Fig.\
\ref{phase_j1j2j3:fig} the essential qualitative features of the spin chain
phase diagram for a generic $J_3=\alpha J_2$ with $0<\alpha\le 1/2$.
The model has a spin liquid XY-like phase at small $J_2$ and $J_z$,
which corresponds, in the fermion language, to a spinless Luttinger liquid
characterized by a certain Luttinger exponent $K$.
(See \ref{details:sec} for more details on this discussion.)
At a given universal value of the Luttinger exponent ($K=1/2$), the
Luttinger liquid phase becomes unstable -- because of {\em umklapp
processes\/} -- against two different (gapped) ordered phases, depending on
the sign of the effective coupling of the umklapp term:
a N\'eel phase with $\uparrow\downarrow\uparrow\downarrow$ LRO,
for large $J_z$ and small $J_2$, and a dimer phase, for larger
$J_2$.
Both phases have a {\em gap\/} in the excitation spectrum, and a
{\em doubly degenerate\/} ground state which {\em breaks translational
symmetry\/}.\cite{Hal_82,Nomura}
These two ordered phases are separated by a critical line of continuously
varying exponent, labelled `PM' in Fig.\ \ref{phase_j1j2j3:fig}, along which
the effective coupling of the umklapp term vanishes and the system
behaves as a Luttinger model with a Luttinger exponent $1/8<K<1/2$.
Beyond the point `M' in Fig.\ \ref{phase_j1j2j3:fig}, the nature of the line
changes from non-universal to first order.
For even larger values of $J_2$ the other ordered phase,
with spins acquiring $\uparrow\uparrow\downarrow\downarrow$ LRO and
a fourfold degenerate ground state, wins over the dimer phase.
This is the only feature of Figs.\ \ref{phase_j1j2:fig} and
\ref{phase_j1j2j3:fig} which is qualitatively new with
respect to the phase diagram of the $J-J_2$--model
(Eq.\ \ref{HAL_MOD:eqn}).\cite{DIFF_MOD:nota}

The nature of the line separating the $\uparrow\uparrow\downarrow\downarrow$
phase from the dimer phase is an open issue.
Previous studies of the $K_3$--model \cite{Mazzeo} and of the spin
chain \cite{pepin} found exponents which appeared to be compatible
with the 2D-Ising universality class. Recently, a transfer matrix study
of a $2D$ model closely related to the $K_4$--model has found exponents
which are incompatible with Ising.\cite{bastia}
A definite answer to the nature of this transition, possibly connected to
the presence or absence of the multicritical point M in the phase diagram,
\cite{bastia} needs further study.
In spite of this uncertainty, we will continue to refer to this line, for
convenience, as ``Ising''.

A second open issue concerns the region of the phase diagram
where the KT line and the ``Ising'' line seem to approach each other.
A relevant question, which we have not been able to answer, is whether the KT
and the ``Ising'' lines actually merge, and, if so, what is the nature of the
resulting line.

\subsection{Phase diagram of the modified BCSOS models}
\label{BCSOS_phase_dia:sec}

The translation of Fig.\ \ref{phase_j1j2:fig} into a temperature phase
diagram for $K_3$--model, using Eqs.\ \ref{COUP_RELAT:eqn}, is shown in
Fig.\ \ref{phase_k3:fig}.\cite{PHASE_DIAG:nota}
The generic phase diagram of our modified BCSOS model, in the $(T,K_{2x})$
plane for given values of $K_3$ and $K_4$, is qualitatively sketched in
Fig.\ \ref{phase_k3k4:fig}.\cite{k3_k3k4_diff:nota}
Four phases are found in a region of parameters relevant to the
unreconstructed and $(2\times 1)$ MR reconstructed case.
At very high temperatures, there is a rough phase.
It corresponds, in the spin problem, to the region close to XY-model point
($J_z=J_2=J_3=0$) in which spin-spin correlation functions decay as power laws
at large distances (the Luttinger liquid or Gaussian model, see
subsection \ref{details:sec}).
A large-distance uniform term of the type $-K/(2\pi^2n^2)$ in the spin-spin
correlation function $\langle S^z_0 S^z_n\rangle$ -- see Eq.\
\ref{density_corr:eqn} -- implies a logarithmic divergent
height-height correlation function
$G(n)=\langle [h^{(0)}_n-h^{(0)}_0]^2\rangle$, signalling a rough phase.
Indeed, using Eq.\ \ref{RELAT_H_SPIN:eqn} and translational invariance of the
spin-spin correlation function, one verifies that
\begin{eqnarray} \label{hh_corr:eqn}
G(n) &=& \langle [h^{(0)}_n - h^{(0)}_0]^2 \rangle \nonumber \\
     &=& 4 \sum_{i,j=0}^{n-1} \langle S^z_i S^z_j \rangle
     = n + 8\sum_{i=1}^n (n-i) \langle S^z_0 S^z_n \rangle \nonumber \\
     &=& \frac{4K}{\pi^2} \ln{(n)} \,+\, \cdots, \hspace{10mm} n\to\infty \;.
\end{eqnarray}
At low temperatures, corresponding to large values of $J_z/J$ and/or
$J_2/J$ in the spin-chain problem, a $(1\times 1)$ and a $(2\times 1)$
ordered phase are present for $K_{2x}>0$ and $K_{2x}<0$, respectively.
The $(1\times 1)$ and $(2\times 1)$ ordered phases correspond, respectively,
to $\uparrow\downarrow\uparrow\downarrow$ and
$\uparrow\uparrow\downarrow\downarrow$ LRO for the spins
(see Fig.\ \ref{gs:fig}).
The other phase appearing in Figs.\ \ref{phase_k3:fig} and \ref{phase_k3k4:fig}
is a disordered flat (DF) phase.
It corresponds, in the spin language, to the dimer phase (see Section
\ref{dimer:sec} for a more extensive discussion).
The transition line from the $(2\times 1)$ reconstructed phase to the
DF phase is labelled as ``Ising'', in spite of the fact that its nature
is not completely assessed (see previous section).
The critical line separating the unreconstructed phase from the disordered
flat phase has variable exponents: it is the
{\em preroughening line\/}.\cite{MdN_Rom}
The parameter $K$ appearing in Eq.\ \ref{hh_corr:eqn} is the Luttinger
exponent. In the rough phase $K>1/2$. Along the preroughening line
correlation functions still behave as power laws with exponents
related to $K$; Eq.\ \ref{hh_corr:eqn} is still valid, with $1/8<K<1/2$.

\subsection{Spinless Luttinger liquid and the variable exponent line}
\label{details:sec}

We now discuss in more detail how to extract, using
standard techniques of one-dimensional systems, a qualitative phase diagram
for our spin chain model and some useful information about the variable
exponent line.
The reader not interested in technical details might jump directly to the
next section, where the surface interpretation of the dimer spin phase
is discussed.

The starting point is the spinless fermion model in Eq.\ \ref{FERMION_H_1:eqn}.
The low-energy physics of such a model, as long as the interactions are not
too strong, can be conveniently analyzed by going to the continuum limit,
$a\to 0$ with $Na=L$ fixed.
One linearizes the fermionic band around the two Fermi points at $\pm k_F$,
and introduces a right ($p=+$) and left ($p=-$) branch of fermions, with
fields $\psi_p(x)$.\cite{Hal_81}
All the interactions processes in which particles are scattered in the
vicinity of the Fermi points are then classified in the so-called
``g-ology'' scheme.\cite{Solyom}
The resulting {\em continuum\/} fermionic model ${\cal H}_F$ turns out
to be a sum of two terms
\begin{equation} \label{fermion_continuum:eqn}
{\cal H}_F = {\cal H}_{\rm Luttinger} + {\cal H}_{\rm umklapp} \;,
\end{equation}
where ${\cal H}_{\rm Luttinger}$ is a spinless Luttinger model,\cite{Hal_81}
\begin{eqnarray} \label{Luttinger:eqn}
{\cal H}_{\rm Luttinger} &=& v_F \sum_{p=\pm} \int_0^L \! dx \,
: \psi^{\dagger}_p(x) [-ip\nabla-k_F] \psi_p(x) : \nonumber \\
&& \hspace{5mm} + \sum_{p,p'=\pm} [g_4\delta_{p',p}+g_2\delta_{p',-p}]
\int_0^L \! dx \, :\rho_p(x) \rho_{p'}(x): \;,
\end{eqnarray}
and ${\cal H}_{\rm umklapp}$ represents the crucial
{\em umklapp processes\/}, i.e., processes where two fermions
are scattered from the vicinity of one Fermi point to the opposite
one,\cite{Solyom}
\begin{equation} \label{umklapp:eqn}
{\cal H}_{\rm umklapp} =
g_3 \int_0^L \! dx \, [ :\psi^{\dagger}_+(x) \psi_-(x) \psi^{\dagger}_+(x)
\psi_-(x): + H.c.] \;.
\end{equation}
(Umklapp processes would not conserve the momentum for a general filling: at
half-filling, however, momentum conservation is fulfilled
modulo a reciprocal lattice vector, $G=4k_F=2\pi$.)
Here $v_F=2t=J$ is the Fermi velocity, and
$\rho_p(x)=:\psi_p^{\dagger}(x)\psi_p(x):$ is the density operator
for the $p$-branch of fermions. (The $:\cdots :$ stands for a normal ordering
procedure, as explained in \cite{Hal_81}.)
Neglecting lattice renormalization effects we have, for the Luttinger couplings
$g_4=v(0)=(J_z+J_2+J_3)$ and $g_2=v(0)-v(\pi)=2(J_z+J_3)$, whereas the umklapp
coupling reads $g_3=v(\pi)=(-J_z+J_2-J_3)$.
We stress the important point that {\em the sign of umklapp coupling $g_3$
results from a competition of $J_z$ and $J_2$\/}. We will see that this fact
is crucial to the existence of a line with variable exponents.

The final step is to {\em bosonize\/} the hamiltonian in
Eq.\ \ref{fermion_continuum:eqn}.
This is achieved by introducing a bosonic representation
of the fermionic fields \cite{Hal_81,Solyom}
\begin{equation} \label{fermi_bose:eqn}
\psi_p(x) \,=\, \frac{1}{\sqrt{2\pi\alpha}} \, \eta_p \, e^{ip k_F x}\,
e^{ip \phi_p(x)} \;,
\end{equation}
where $\alpha$ is a short-distance cut-off, and
$\eta_p=\eta_p^{\dagger}$ are Majorana fermions ($\eta^2_p=1$) ensuring
correct anticommutation properties among right and left-moving fermions.
The field $\phi_p(x)$ is related to the fermion density as follows
\begin{equation}
\rho_p(x) \,=\, : \psi_p^{\dagger}(x) \psi_p(x) :
\,=\, \frac{1}{2\pi} \nabla \phi_p(x) \;,
\end{equation}
and is expressed in terms of standard boson operators $b_p(q)$ as:
%(i.e., $[b_p(q),b^{\dagger}_{p'}(q')]=\delta_{p,p'}\delta_{q,q'}$) as:
%
\[ \phi_p(x) = \sum_{q>0} e^{-\alpha q/2}\left( \frac{2\pi}{Lq} \right)^{1/2}
\left[ e^{-ipqx} b_p^{\dagger}(q) + H.c. \right] \;. \]
(Here $q=(2\pi/L)n$, with $n$ integer.)
The continuum model in Eq.\ \ref{fermion_continuum:eqn} translates, in bosonic
variables, into a quantum sine-Gordon problem \cite{Hal_81,Nomura}
\begin{equation}
{\cal H}_{SG} = \frac{v_S}{2} \int_0^L \! dx \,
[K \Pi^2 + \frac{1}{K} (\nabla \Phi )^2] + \frac{V}{(2\pi\alpha)^2}
\int_0^L \! dx \, \cos{(\sqrt{16\pi}\Phi)} \;,
\end{equation}
where we have introduced the canonical field
$\Phi(x)=(\phi_++\phi_-)/\sqrt{4\pi}$ and
its conjugate momentum $\Pi(x)=-\nabla (\phi_+-\phi_-)/\sqrt{4\pi}$.
The Luttinger model, ${\cal H}_{\rm Luttinger}$, is equivalent to free bosons
i.e., to a massless Klein-Gordon (or Gaussian) problem ($V=0$).\cite{Hal_81}
$K$ is the crucial parameter governing the low-energy physics of the problem,
$v_S$ being the velocity of the sound-like gapless excitations.\cite{Hal_81}
In weak-coupling we have $v_S/K=J[1+ (3J_z+J_2+3J_3)/\pi J +\cdots]$, and
$v_SK =J[1- (J_z-J_2+J_3)/\pi J +\cdots]$.
The umklapp term in Eq.\ \ref{umklapp:eqn}, rewritten using
Eq.\ \ref{fermi_bose:eqn}, gives rise to the cosine term, with
$V=-2g_3=2(J_z-J_2+J_3)+\cdots$.
The renormalization group equations for the sine-Gordon problem are well
known,\cite{Giamarchi,Nomura} and have the Kosterlitz-Thouless form
\begin{eqnarray}
\frac{dK}{dl} &=& -\tilde{V}^2(l) \nonumber \\
\frac{d\tilde{V}}{dl} &=& 2[1-2K(l)]\tilde{V}(l) \;,
\end{eqnarray}
with $\tilde{V}$ simply proportional to $V$.
$V=0$ with $K>1/2$ is a line of stable fixed points which represent
the Luttinger (or Gaussian) model.
$V=0$ with $K<1/2$, on the contrary, is a line of {\em unstable fixed
points\/}: the smallest $V\ne 0$ will grow upon renormalization if $K<1/2$,
the system will ``go to strong coupling'' and develop a gap in the
excitation spectrum.

At the XY point, $v_S=v_F=J$ and $K=1$.
For small values of the couplings $J_z,J_2,J_3$ the exponent $K$ is
larger than $1/2$ and the umklapp term is irrelevant, $V\to 0$.
This region corresponds to a {\em spin liquid\/}.
The large distance behaviour of the correlation functions is characterized
by power laws with exponents related to $K$.
For instance, spin-spin correlations like $\langle S^z_0 S^z_n \rangle$ are
related (recall that $S^z_i=n_i-1/2$) to density-density correlations of the
spinless fermions $\langle n(0) n(x) \rangle$.
The density operator has a continuum limit expression of the type
\begin{eqnarray} \label{density_op:eqn}
n(x) &\sim& (\rho_+(x) + \rho_-(x)) \;+\; [\psi_+^{\dagger}(x) \psi_-(x) +
H.c.]
\nonumber \\
&=& \frac{1}{\sqrt{\pi}} \nabla \Phi(x) + \frac{1}{\pi\alpha}
\sin{[\sqrt{4\pi}\Phi(x)+2k_F x]} \;.
\end{eqnarray}
Using the fact that correlation functions of the bosonic field are simple
to calculate for the Gaussian model ($V=0$), i.e.,
\begin{equation}
G(x) = \langle \Phi(x) \Phi(0) - \Phi^2(0) \rangle_{V=0} = \frac{1}{4\pi}
\ln{ \frac{\alpha^2}{\alpha^2+x^2} } \;,
\end{equation}
\begin{equation}
\langle e^{i\gamma\Phi(x)} e^{-i\gamma\Phi(0)} \rangle_{V=0}
= e^{\gamma^2 G(x)}
= \left[ \frac{\alpha^2}{\alpha^2+x^2}  \right]^{\gamma^2/(4\pi)} \;,
\end{equation}
it is simple to show that
\begin{equation} \label{density_corr:eqn}
\langle n(0) n(x) \rangle = -\frac{K}{2\pi^2x^2} +
A \frac{ \cos{2k_Fx} }{x^{2K}} + \cdots \;,
\end{equation}
$A$ being a non-universal constant.

Increasing the values of the couplings, the spin liquid phase becomes
unstable, at $K=1/2$, against two different gapped phases, depending
on the sign of the umklapp term $V$. For $V>0$ (large values of $J_z$)
the strong coupling fixed point is characterized by a field $\Phi$ which
is frozen at a value such that $\cos{\sqrt{16\pi}\Phi}=-1$,
i.e., $\sqrt{4\pi}\Phi=\pi/2$.
It is then clear that density-density correlations acquire LRO, since from
Eq.\ \ref{density_op:eqn} we get
\[ \langle n(0) n(x) \rangle \sim \sin{(\pi/2+2k_Fx)} = \cos{(2k_Fx)} \]
signalling a site-centered $2k_F$ charge density wave (CDW).
In the spin language this corresponds to a N\'eel phase with
$\uparrow\downarrow\uparrow\downarrow$ LRO.
For $V<0$, on the contrary, the strong coupling fixed point is characterized
by a field $\Phi$ which is frozen at the value $\Phi=0$ (or $2\pi$).
To guess what correlation functions acquire LRO, notice
that the canonical transformation
$\psi_p \to e^{-ip\pi/4} \psi_p$ changes the sign of the umklapp term in
Eq.\ \ref{umklapp:eqn},\cite{Hal_82}
(in boson language this corresponds to
$\sqrt{4\pi}\Phi \to \sqrt{4\pi}\Phi+\pi/2$).
Knowing that the $2k_F$-component of the density operator acquires LRO in
the N\'eel phase ($V>0$), we immediately conclude that the
operator having LRO for $V<0$ reads
\begin{equation} \label{dimer_op:eqn}
i [\psi_+^{\dagger}(x) \psi_-(x) - H.c.]
\sim  \frac{1}{\pi\alpha}
\cos{[\sqrt{4\pi}\Phi(x)+2k_F x]} \;.
\end{equation}
An operator whose continuum limit $2k_F$-component is given
by Eq.\ \ref{dimer_op:eqn} is readily found to be the bond kinetic energy
$(c^{\dagger}_{i} c_{i+1} + c^{\dagger}_{i+1} c_{i})$.
The strong coupling phase obtained for $V<0$ is therefore
a {\em bond-centered charge density wave\/}, to be contrasted to the
{\em site-centered\/} CDW obtained for $V>0$.
In spin language, this bond-centered CDW is a {\em spin dimer\/} phase.

Separating the N\'eel ($V>0$) from the dimer phase ($V<0$) is the line
of unstable fixed points ($V=0$ with $K<1/2$) mentioned above. Along
this line (`PM' in Fig.\ \ref{phase_j1j2j3:fig}), the system behaves
as an effective Luttinger (or Gaussian) model with $1/8<K<1/2$.
If $K<1/8$, cosine terms of the type $V'\cos{(2\sqrt{16\pi}\Phi)}$
-- formally coming from higher order umklapp processes involving
four-particle scattering -- become relevant and open up a gap. The
nature of the transition line changes to first order.
Correlation functions behave as power laws along the line PM.
Density-density correlations, for instance, are still given by
Eq.\ \ref{density_corr:eqn}.
All the critical exponents along the $V=0$--line are known in terms of $K$.
The correlation function exponent follows directly from
Eq.\ \ref{density_corr:eqn}, i.e., $\eta=2K$.
The gap between the ground state and the first excited state goes
like\cite{Solyom}
\[ \Delta = \frac{1}{\xi} \sim |V|^{1/(2-4K)} \;, \]
implying a correlation-length exponent $\nu=1/(2-4K)$.
The order parameter exponent is given by $\beta=\nu K$. The specific
heat exponent is $\alpha=2-2\nu=(2-8K)/(2-4K)$.

\section{The spin dimer phase and its surface interpretation}
\label{dimer:sec}

In the spin dimer phase, ordinary spin--spin correlations decay exponentially
to zero, but four--spin correlation functions of the type
$\langle ({\vec S}_i\cdot {\vec S}_{i+1})
({\vec S}_j\cdot {\vec S}_{j+1})\rangle$ acquire LRO.\cite{Hal_82}
More specifically, everywhere inside the dimer phase in
Figs.\ \ref{phase_j1j2:fig}-\ref{phase_j1j2j3:fig}, one has:
\begin{eqnarray}
S^{zz}_j &\,=\,& \langle \, S^z_0 S^z_j \, \rangle \to 0
\hspace{10mm} j\rightarrow \infty \, \nonumber \\
S^{\rm dim}_j &\,=\,&
\langle\, (S^z_0 S^z_{1})\, (S^z_j S^z_{j+1})\, \rangle \approx
A + B (-1)^{j} \hspace{10mm} j\rightarrow \infty \;.
\end{eqnarray}
This is illustrated in Fig.\ \ref{sq:fig}, where we show the size
dependence of various static structure factors at the point
($J_z=3J$,$J_2=2.4J$,$J_3=0$).
These values are obtained from exact diagonalizations of chains up to
$28$ sites.
The solid squares represent the dimer static structure factor at $q=\pi$
\begin{equation}
S^{\rm dim}(q=\pi) = \sum_j e^{i\pi j} S^{\rm dim}_j \;,
\end{equation}
whereas the open squares and the stars represent, respectively, the $\pi/2$ and
$\pi$ component of the ordinary spin-spin structure factor.
Clearly, $S^{\rm dim}(q=\pi)$ diverges {\em linearly\/} with the length of the
chain (see inset of Fig.\ \ref{sq:fig}), implying long-ranged oscillations
of the corresponding correlation function, whereas the usual spin-spin
structure factor is finite.

To illustrate in more detail some of the physics of this disordered spin
state, and its translation into the surface language, we consider a
representative dimer phase point.
As it happens, there is a special point in the phase diagram of
the $J-J_2$--model (Eq.\ \ref{HAL_MOD:eqn} with $J_z=J=2J_2$), where the
twofold degenerate exact ground state is exactly known,\cite{Mad_Gho} and
extremely simple: it is just a product of spin singlets.
Explicitly, for any finite (even) size $N$ the two ground states,
which turn into one another by translation of a lattice spacing, are
\begin{eqnarray} \label{DIMERS:eqn}
|\Psi_1\rangle &=& |1 2\rangle |3 4\rangle \cdots |N-1 N\rangle \nonumber \\
|\Psi_2\rangle &=& |2 3\rangle |4 5\rangle \cdots |N-2 N-1\rangle
|N 1\rangle \;.
\end{eqnarray}
Here $|i j\rangle=|\uparrow\downarrow-\downarrow\uparrow\rangle/\sqrt{2}$
denotes a singlet between sites $i$ and $j$.
Eq.\ \ref{Bond_CDW:eqn} is just the spinless fermion translation of $\Psi_2$.
Some of the properties of these prototype dimer states which we are going
to illustrate have been first discussed, in connection to the DF
phase problem, in Ref.\ \cite{MdN_Rom}.
Obvious properties of $|\Psi_1\rangle$ are, for instance, that spin--spin
correlations are extremely short ranged,
\begin{eqnarray} \label{PROP1:eqn}
\langle\Psi_1| S^z_j | \Psi_1 \rangle &=& 0 \hspace{10mm} \forall j \nonumber
\\
\langle\Psi_1| S^z_{i} S^z_{j} |\Psi_1\rangle &=& 0  \hspace{10mm} |i-j|>1 \;,
\end{eqnarray}
and that translational invariance is spontaneously broken,
\begin{eqnarray} \label{PROP2:eqn}
\langle\Psi_1| S^z_{2j-1} S^z_{2j} |\Psi_1\rangle &=& -1/4    \nonumber \\
\langle\Psi_1| S^z_{2j} S^z_{2j+1} |\Psi_1\rangle &=& 0        \;.
\end{eqnarray}
In spite of this order parameter, such states are clearly
spin-{\em disordered\/}.
$(1\times 1)$ order for the surface, for instance, translates into
N\'eel LRO for the spin chain (see Fig.\ \ref{gs:fig}),
whereas a dimer state has only short range spin--spin correlations.
To see why they describe a {\em flat\/} surface, consider expanding the
product of singlets in Eq.\ \ref{DIMERS:eqn} for $|\Psi_1\rangle$, say.
One obtains the sum of $2^{(N/2)}$ spin configurations, one of which
will be of the typical form
\begin{equation}
(\uparrow \downarrow) \, (\uparrow \downarrow) \;
(\downarrow \uparrow) \, (\downarrow \uparrow) (\downarrow \uparrow) \;
(\uparrow \downarrow) \, (\uparrow \downarrow)  \cdots \;.
\end{equation}
Here we have taken the $(\uparrow \downarrow)$ part of the singlet for the
first two pair of sites, the $(\downarrow \uparrow)$ part of the singlet for
the next three pairs of sites, and so on.
A {\em down\/} $(2\times 1)$ step
(i.e., a pair of neighboring down spins, see Fig.\ \ref{defects:fig})
is obtained each time a $(\downarrow \uparrow)$ pair follows immediately
after a $(\uparrow \downarrow)$ one, and, vice-versa, an {\em up\/}
$(2\times 1)$ step (a pair of neighboring up spins) results from a
$(\uparrow \downarrow)$ pair following a $(\downarrow \uparrow)$ one.
In between steps, there are regions with N\'eel type of order
(unreconstructed regions in the surface language).
Clearly, there is no way of having two up steps (or two down steps)
following each other: a step up is followed necessarily by a step down
and vice-versa. The surface is therefore {\em flat}.\cite{MdN_Rom}

In the dimer phase there are characteristic correlations between steps that
are worth stressing.
An up (down) $(2\times 1)$ step ending at site $j$ is ``measured''
(see previous discussion) by the spin operator
\[ {\rm Step}^{\pm}_j = (S^z_{j-1}\pm 1/2) (S^z_{j}\pm 1/2) \;, \]
counting, respectively, $\uparrow\uparrow$ (${\rm Step}^+$) and
$\downarrow\downarrow$ (${\rm Step}^-$) combinations at sites $(j-1,j)$.
An operator counting a step, irrespective of its being up or down, is given by
\begin{equation} \label{step_def:eqn}
{\rm Step}_j = {\rm Step}^{+}_j + {\rm Step}^{-}_j =
2 (S^z_{j-1} S^z_{j} + 1/4) \;.
\end{equation}
One can easily work out correlation functions for such step operators in the
representative dimers states. For odd $j$, for instance, one finds:
\[
\langle \Psi_1| {\rm Step}^{+}_j {\rm Step}^{+}_{j+n} | \Psi_1\rangle \,=\,
\left\{ \begin{array}{ll}
          \frac{1}{4}  & \mbox{if $n=0$} \\
          \frac{1}{16} & \mbox{if $n>2$ and even} \\
                0      & \mbox{otherwise}
       \end{array}     \right. \;, \]
\[
\langle \Psi_1| {\rm Step}^{+}_j {\rm Step}^{-}_{j+n} | \Psi_1\rangle \,=\,
\left\{ \begin{array}{ll}
          \frac{1}{8}  & \mbox{if $n=2$} \\
          \frac{1}{16} & \mbox{if $n>2$ and even} \\
                0      & \mbox{otherwise}
       \end{array}     \right. \;, \]
\begin{equation} \label{step_step_corr:eqn}
\langle \Psi_1| {\rm Step}_j {\rm Step}_{j+n} | \Psi_1\rangle \,=\,
\left\{ \begin{array}{ll}
          \frac{1}{2}  & \mbox{if $n=0$} \\
          \frac{1}{4}  & \mbox{if $n\ge 2$ and even} \\
                0      & \mbox{otherwise}
       \end{array}     \right. \;.
\end{equation}
Similar results apply to $\Psi_2$ for the case of even $j$.
It is interesting to see how closely a point inside the dimer phase
of Fig.\ \ref{phase_j1j2:fig} resembles such an ideal
scenario.\cite{step_3x1:nota}
Fig.\ \ref{step:fig} show step-step correlations
$\langle {\rm Step}_j {\rm Step}_{j+n} \rangle$
obtained from exact diagonalization of a chain of $28$ sites,
for a point inside (a) the dimer phase ($J_z=3.0J$,$J_2=2.4J$,$J_3=0$),
and (b) the N\'eel phase ($J_z=3.0J$,$J_2=J_3=0$).
In the N\'eel phase, $\uparrow\uparrow$ and $\downarrow\downarrow$
steps are bound in pairs, and the correlation function decays
exponentially to the square of the step-concentration,
shown by a dashed line in Fig.\ \ref{step:fig} (b).
The relevant defect is therefore the domain wall denoted by
$\epsilon^*_{2\times 1}$ in Fig.\ \ref{defects:fig}.
In the dimer phase, on the contrary, $\uparrow\uparrow$ and
$\downarrow\downarrow$ steps are unbound, and free to move in a fluid-like
manner, but their correlation function displays long-ranged oscillations
with period $\pi$. In other words, the fluid of (roughly) alternating
up and down steps has the feature that steps prefer to stay at an
even distance from each other.
In the neighboring $(2\times 1)$ phase, this fluid of $2\times 1$ steps
solidifies into an ordered structure of the type
$\uparrow\uparrow\downarrow\downarrow$.
We stress the fact that the oscillations displayed in
Fig.\ \ref{step:fig} (a) are not due to $(2\times 1)$ order; the point
considered is, as demonstrated in Fig.\ \ref{sq:fig}, disordered.

Step-step correlations of the type shown are simple manifestations
of the spontaneous breaking of translational invariance.
Similar (and related) effects can be seen in other properties of the
disordered surface.
Suppose we want to count, in the surface terminology, the difference in the
number of white and black local maxima in the surface.
We restrict first our considerations to sites which are local maxima when
considered in the x-direction only.
In the spin language, a local ``maximum'' at site $j$ occurs whenever the
site $j-1$ has spin $\uparrow$ and the site $j$ has spin $\downarrow$.
An operator which ``counts'' the maximum at $j$ is therefore
$(S^z_{j-1}+1/2)(1/2-S^z_j)$.
The difference between white (even $j$) and black (odd $j$) maxima is therefore
measured by the order parameter
\[
P^{\rm (spin)}_{BW}=(2/N)\sum_j e^{i\pi j} (S^z_{j-1}+1/2)(1/2-S^z_j) \;.
\]
$P^{\rm (spin)}_{BW}$ is odd under translation.
Its value is $1$ on the N\'eel state
$|\uparrow\downarrow\uparrow\downarrow\cdots\rangle$, and $-1$ on the other
N\'eel state $|\downarrow\uparrow\downarrow\uparrow\cdots\rangle$.
Quite generally, it is different from zero in the whole N\'eel phase of the
spin phase diagram.
Consider now the value of $P^{\rm (spin)}_{BW}$ on the dimer state
$|\Psi_1\rangle$. Using the elementary results in
Eq.\ \ref{PROP1:eqn}-\ref{PROP2:eqn}, we arrive at
\begin{equation} \label{PBW_DIMER:eqn}
\langle\Psi_1| P^{\rm (spin)}_{BW} |\Psi_1\rangle
= -\frac{2}{N} \sum_j e^{i\pi j}
\langle\Psi_1| S^z_{j-1} S^z_{j} |\Psi_1\rangle = \frac{1}{4} \;.
\end{equation}
Similarly, $\langle\Psi_2| P^{\rm (spin)}_{BW} |\Psi_2\rangle = -1/4$.
Therefore, the implication of the dimer scenario, with its spontaneous
breaking of translational symmetry, is that, on the disordered flat surface,
one of the two sublattices tends to dominate in the local maxima.

One can check this prediction by Monte Carlo simulations of the original
classical models.
In the next section we will present the results of our simulations for
the $K_3$ and the $K_4$--model.
The results strongly support the dimer phase scenario.

\section{Monte Carlo results and discussion}
\label{mc:sec}

We have performed classical Monte Carlo simulations of
the $K_3$-- and $K_4$--model in the DF phase.
We have measured, to start with, standard quantities like the square mean width
of the surface, $\delta h^2$,
\begin{equation} \label{dh2:eqn}
\delta h^2 \,=\, \langle \frac{1}{8N_c^2} \sum_{{\bf r},{\bf r}'}
  (h_{\bf r}-h_{{\bf r}'})^2 \rangle \;,
\end{equation}
the $(1\times 1)$ order parameter, $P_{1\times 1}$, and the $(2\times 1)$
reconstruction one, $P_{2\times 1}$,
\begin{eqnarray} \label{P1x1:eqn}
P_{1\times 1} &=& \langle \frac{1}{N_c} \sum_{\bf r}
h_{\bf r} e^{i{\bf G} \cdot {\bf r}} \rangle =
\langle \frac{1}{N_c} \sum_{{\bf r}\in W}
[h_{\bf r} - h_{{\bf r}+{\bf b}}] \rangle \;, \nonumber \\
P_{2\times 1} &=& \langle \frac{1}{2N_c} \sum_{\bf r}
h_{\bf r} e^{i{\bf G} \cdot {\bf r}/2} \rangle \;.
\end{eqnarray}
Here $N_c$ is the number of cells in each sublattice (i.e., $2N_c$ is the
number of atoms), and ${\bf G}=(2\pi/a_x) {\hat {\bf x}}$.
The square mean width $\delta h^2$ diverges logarithmically in the rough
phase as the size of the sample $L\rightarrow \infty$
\[ \delta h^2 \approx K(T) \log{L}  \;, \]
with a coefficient $K(T)$ larger than a (universal) minimum value
$K(T_R)=1/\pi^2$ attained at the roughening temperature.
$P_{1\times 1}$ is different from zero only in the unreconstructed region of
the
phase diagram and goes to zero at the preroughening line.
$P_{2\times 1}$ is different from zero in the reconstructed region of the
phase diagram and goes to zero at the ``Ising'' line.
Clearly, the DF phase has $P_{1\times 1}=0$, $P_{2\times 1}=0$,
and $\delta h^2<\infty$.
On the basis of the spin mapping and of the dimer phase scenario we
expect, however, that some form of order will be present:
one should be able to tell which of the two sublattices (W or B) prevails
in the top layer.
A way of testing this is to define the ``local peak'' operator
\begin{equation}
O_{\bf r} = \frac{1}{16} \prod_{i=1}^4 [\Delta h_{{\bf r},i} + 1]
\end{equation}
where $\Delta h_{{\bf r},i}=h_{\bf r} - h_{{\bf r}+{\bf b}_i}$ and
${\bf b}_i$ with $i=1,\cdots,4$ are the vectors connecting a chosen site to
its four nearest neighbors (belonging to the opposite sublattice).
$O_{\bf r}$ takes the value 1 for the atoms lying above all their neighbors,
and zero otherwise.
Summing over all the sites with a phase factor $1$ for the W sites and $-1$
for the B ones, we get a quantity measuring which sublattice prevails in the
top layer,
\begin{equation} \label{pbw:eqn}
P_{BW} = \langle \frac{1}{N_c} \sum_{\bf r} e^{i{\bf G}\cdot {\bf r}} O_{\bf r}
\rangle \;.
\end{equation}
As defined, $P_{BW}$ is normalized to 1 on the unreconstructed ground states,
and to 1/2 on the reconstructed $(2\times 1)$ ground states.\cite{PBW_PEO:nota}
Our expectation is that {\em $P_{BW}$ is different from zero in the disordered
flat phase\/}, and vanishes in the rough region and on the {\em preroughening
line\/}. (See Fig.\ \ref{pbw:fig}.)

A classical grand-canonical single-move Monte Carlo code has been set up and
used for lattices of linear size $L=N_x=N_y$ up to 100.
Starting from a disordered surface, we randomly add or remove particles,
making sure that the BCSOS constraint is fulfilled at each step,
and accept moves according to the standard Metropolis algorithm.
The configurations resulting from consecutive sweeps of the lattice
($2L^2$ attempted moves) are quite correlated, so that independent values for
the various averages are obtained as a result of a sufficiently large number
of Monte Carlo sweeps.
It is on the basis of such ``independent measurements'' that statistical
errors are estimated.
Typically 20 to 50 such measurements are performed, each of which consists
of $10^5-10^6$ sweeps, after a suitable equilibration of the system.

For the $K_3$--model, we used the parameters of Mazzeo {\em et al.},
roughly chosen to fit the glue model results of Ercolessi
{\em et al.} \cite{Ercolessi} for gold:
$K_{2x}/K_{2y}=-0.51$, $K_3/K_{2y}=0.22$ (i.e., ${\cal K}=-2.3$).
An Ising type deconstruction transition has been reported to take place at
$T_D \approx 2.90 K_{2y}$, while a Kosterlitz-Thouless roughening transition
has been found at $T_R \approx 3.09 K_{2y}$.\cite{Mazzeo}
We have performed a careful finite-size scaling analysis of the different
order parameters at the intermediate temperature $T=3.0K_{2y}$.
The surface is still flat at this temperature, as demonstrated in
Fig.\ \ref{auk3_dh2:fig}, showing that $\delta h^2$ versus $\log{L}$ stays
definitely below the universal critical slope $K(T_R)=1/\pi^2$,
which implies that $\delta h^2$ will eventually saturate to a constant as
$L\rightarrow \infty$.
Fig.\ \ref{auk3:fig} (a) shows the results obtained for $P_{2\times 1}$
(solid circles), and $P_{BW}$ (diamonds).
The squares denote a further order parameter used by
Mazzeo {\em et al.} \cite{Mazzeo},
\begin{equation} \label{pbw2x1:eqn}
P^{(2\times 1)}_{BW} = \langle \frac{1}{4N_c} \;
\left[ \sum_{{\bf r}\in W} |S_{\bf r}| - \sum_{{\bf r}\in B}|S_{\bf r}|
\right] \rangle \;,
\end{equation}
where the classical ``spin'' variables $S_{\bf r}$ are defined in terms of
the nearest neighbor height differences as
\begin{equation}
S_{\bf r} = \sum_{i=1}^4 \Delta h_{{\bf r},i}.\\
\end{equation}
$P_{2\times 1}$ vanishes as $L^{-1}$, see inset of Fig.\ \ref{auk3:fig} (a),
confirming that $T=3.0K_{2y}$ is above the deconstruction temperature $T_D$,
in agreement with Ref.\ \cite{Mazzeo}.
Both $P_{BW}$ and $P^{(2\times 1)}_{BW}$ decrease, instead, much slower
than $L^{-1}$.
Fig.\ \ref{auk3:fig} (b) shows the logarithm of $P_{BW}$ versus $\log L$.
The data for small sizes ($L$ up to 48) can be fit with a power law
$L^{-0.37}$.
For larger values of $L$, a crossover is seen to what is most probably an
exponential convergence to {\em a non zero limit} for $P_{BW}$.
In other words, systems up to $L=48$ are still smaller than the actual
value of a correlation length $\xi_{BW}$, so that a fictitious power law
behavior is initially seen.
A similar behavior is also found for $P^{(2\times 1)}_{BW}$.

The corresponding results for a point inside the DF phase
of the $K_4$--model ($K_4/K_{2y}=0.1$, $K_{2x}/K_{2y}=-0.056$,
and $T/K_{2y}=2.3$) are shown in Fig.\ \ref{auk4:fig}.
Entirely similar comments apply to this case.

The point in the $K_4$--model phase diagram to which Fig.\ \ref{auk4:fig}
refers to, is in fact located close to the preroughening line.\cite{pepin}
A typical snapshot of the way this disordered flat surface looks like
at this temperature is shown in Fig.\ \ref{snapshot:fig}.
Strictly speaking we are in a parameter region where the classical
ground state is $(2\times 1)$ MR reconstructed and the most
energetically favored defects are {\em Ising\/} walls (see table 1).
However, Ising walls in their ideal form (see Fig.\ \ref{defects:fig})
are almost totally absent.
What one finds, instead, are extended walls of the Ising type with a width
of arbitrary length.
These are nothing but large $(1\times 1)$ unreconstructed regions lying
between two opposite $(2\times 1)$ steps.
Such $(2\times 1)$ steps, which are the very building blocks of a MR
structure, are now free to move in a fluid-like manner with the only constraint
that an up-step is followed by a down-step.
Occasionally, sequences of up-down $(2\times 1)$ steps gain positional order
by ``solidifying'' in $(2\times 1)$ MR regions which are, however,
always of the same ``color'' (more precisely, black, for the phase
illustrated in Fig.\ \ref{snapshot:fig}).
Overall, the surface seems to have as many black regions as white ones: the
$P_{1\times 1}$ order parameter, which counts precisely the relative abundance
of W and B $(1\times 1)$ elementary cells, is small, and goes to zero in
the thermodynamic limit.
Correlations between steps, however, or, in more elementary terms, the fact
that every $(2\times 1)$ step always ends into a B top atom, result in the
above mentioned feature of the absence of white MR regions, and are
such that $P_{BW}$ turns out to be different from zero, albeit small.
Altogether Fig.\ \ref{snapshot:fig} is a nice illustration of how a
dimer disordered flat phase should look like.

These features should be of some relevance in the context of surface
scattering experiments.
We discuss here the case of He scattering.
In the kinematical approximation, and within a SOS framework, the intensity
of the specular peak (parallel momentum transfer ${\bf Q}\approx 0$) with
perpendicular momentum transfer in the so-called anti-phase configuration
is given by
\begin{equation}
I({\bf Q},q_z=\pi/a_z) \,\propto\,
|\langle \sum_{\bf r} e^{i\pi h_{\bf r}} \alpha_{\bf r} \rangle |^2
\delta_{{\bf Q},0}
\,+\, N_{\rm sites} k_B T \chi({\bf Q})\;,
\end{equation}
where $\alpha_{\bf r}$ is an appropriate ``shadowing factor'' which takes into
account the physical requirement that surface peaks scatter more than
valleys.\cite{Levi,Mazzeo}
The first term is a (Bragg) coherent contribution, proportional to the square
of the order parameter and of the number of sites.
The second contribution, due to incoherent terms, is proportional to the
susceptibility of the order parameter and to the number of sites.
For our BCSOS-type of model, in which $h_{\bf r}$ is even in the W sublattice,
and odd in the B sublattice, one immediately concludes that
$e^{i\pi h_{\bf r}}=e^{i{\bf G}\cdot{\bf r}}$ for any allowed height
configuration.
The coherent part of the specular anti-phase peak
$I^{coh}({\bf Q}=0,q_z=\pi/a_z)$ would therefore be identically zero if all
the surface atoms were to scatter in the same way ($\alpha_{\bf r}=1$ for all
${\bf r}$).
In the opposite assumption that only the local peaks scatter efficiently
($\alpha_{\bf r}=1$ if ${\bf r}$ is a local peak, $\alpha_{\bf r}=0$
otherwise), we obtain that $I^{coh}({\bf Q}=0,q_z=\pi/a_z)$ is exactly
proportional to the square of the $P_{BW}$ order
parameter \cite{Bernasconi}
\begin{equation}
I^{coh}({\bf Q}=0,q_z=\pi/a_z) \,\propto\, N^2_{\rm sites} |P_{BW}|^2 \;.
\end{equation}
Quite generally, for a reasonably large class of choices of
shadowing factors $\alpha_{\bf r}$, the breaking of translational invariance
should guarantee that $I^{coh}({\bf Q}=0,q_z=\pi/a_z)$ is different
from zero (albeit possibly small) in the DF phase considered here.
(More precisely, this is so for all the shadowing factors which can be written
in terms of local operators of the $h_{\bf r}$ variables, whose correlation
function is long-ranged in the DF phase.)

Experimentally, a dimer-type of disordered flat phase would manifest itself
with
a rapid fall of the anti-phase scattering as the critical temperature
is approached, followed by an intermediate temperature region, where the
surface is in the disordered flat phase,
in which a {\em small coherent anti-phase scattering intensity survives}.
This situation is sketched in Fig.\ \ref{pbw:fig}.
By normalizing the scattering intensity to its low-temperature value, a dip
at the critical temperature should be observable even if one considers the
total scattering intensity $I({\bf Q}=0,q_z=\pi/a_z)$, which includes
the incoherent contributions.
(Strictly speaking, these contributions are proportional to the susceptibility
which diverges at the critical temperature as $L^{2-\eta}$, where $L$ is
the size of the system. Since $L^2 = N_{\rm sites}$, the incoherent
contributions will never win over the coherent part ($\propto N^2_{\rm sites}$)
and an overall dip should be observable in the normalized scattering
intensity at the critical temperature.)
Clearly, an important requirement for the dimer scenario, which one should test
experimentally, is that the dominant defects proliferating on the disordered
flat surface are indeed monoatomic, or $(2\times 1)$, steps.
The correlations of such monoatomic steps are, at least in principle, also
accessible by direct imaging techniques, such as fast STM.\cite{Frenken}

We mention here, before ending the section, a particularly simple choice
of shadowing factors, proposed in
Ref.\ \cite{Levi}, which {\em does not\/} involve long-ranged operators:
\[ \alpha_{\bf r} = 2 - \frac{n_{\bf r}}{2} \;, \]
where $n_{\bf r}$ is the number of neighbors of the atom in ${\bf r}$
which are found at a level higher than the atom itself.
This expression {\em linearly\/} interpolates between $\alpha=2$ (local
maximum) and $\alpha=0$ (local minimum), and can can be recast in the form
\[ \alpha_{\bf r} = 1 - \frac{1}{4} \sum_{i=1}^{4}
[ h_{{\bf r}+{\bf b}_i} - h_{\bf r} ] \;, \]
where ${\bf b}_i$ are the vectors connecting site ${\bf r}$ to the four
neighboring sites.
Indeed, by exploiting this linearity, it is very simple to show that such a
choice of $\alpha_{\bf r}$ leads to a $I^{coh}({\bf Q}=0,q_z=\pi/a_z)$ which is
proportional to the square of the $(1\times 1)$ order parameter
$P_{1\times 1}$ (see Eq.\ \ref{P1x1:eqn}),
\begin{equation}
I^{coh}({\bf Q}=0,q_z=\pi/a_z) \,\propto\,
|\langle \sum_{\bf r} e^{i{\bf G} \cdot {\bf r}} h_{\bf r} \rangle |^2
= N_c^2 |P_{1\times 1}|^2 \;,
\end{equation}
and therefore vanishes at and beyond the preroughening line.
Therefore, experimental scattering geometries should be chosen so as to
emphasize peak-atom scattering, if the non monotonic behavior of Fig.\
\ref{pbw:fig}, typical of $P_{BW}$, is to be detected.

\section{Conclusions}
\label{conclusions:sec}

The motivation for the present work was a deeper understanding, based on
well-defined hamiltonians, of the nature of the disordered flat phase
(or phases) occurring in simple lattice models of fcc (110) surfaces.
In particular, for reconstructed surfaces, in the spirit of the distinction
proposed in Ref.\ \cite{Bernasconi} between a DEF (Ising wall dominated) phase
as opposed to a DOF (step dominated) phase, we wished to clarify
which of the two scenarios was at play in simple BCSOS-type models.
The outcome of our study is that neither of those simple prototypes
applies, strictly speaking, to the description of the disordered flat phase
we find, which is, on the contrary, closely related to dimer phase of one
dimensional quantum spin-1/2 systems.

The phase diagram in Fig.\ \ref{phase_k3k4:fig}, very similar to the one
discussed in Ref.\ \cite{bastia}, shows many features that
we believe to be quite robust.
First, a transition between the $(2\times 1)$ MR reconstructed phase and
the DF phase with exponents which appear to be very close to Ising.
(Although the actual nature of the line is an open issue, see end
of Sect.\ \ref{spin_phase_dia:sec}.)
Second, the transition line between the unreconstructed and the DF
phase (preroughening) has variable exponents, as was predicted.\cite{MdN_Rom}
Third, the disordered phase has a non-trivial order parameter $P_{BW}$.
It is quite remarkable that both the microscopic models discussed here
and the cell model of Ref.\ \cite{MdN_92} point in the same direction, to a
disordered flat phase which has a non-vanishing order parameter of the type
of $P_{BW}$. We recognize that such a feature is also present in the
phase diagram of Ref.\ \cite{bastia}.

The obvious open question is whether the disordered phase discussed above
is the only one possible.
In other words, can we build microscopic models were defects
other than $(2\times 1)$-steps play a role and the resulting disordered
flat phase (or phases) has qualitatively different features ?

The discussion has to consider separately, at this stage, the case of
semi--microscopic cell-type models \cite{MdN_92} from that of fully
microscopic surface models.
In the former framework of a coarse-grained description of the system, as
the four state clock-step model of den Nijs \cite{MdN_92}, the stage is
clear and the actors are there: walls and steps.
Since white atoms stay on top in regions where the reconstruction variable
$\theta$ is $0$ or $\pi$, and black atoms do so in regions where $\theta$
is $\pi/2$ or $3\pi/2$, the $P_{BW}$ order parameter has to be non-zero in
the disordered flat phase of this model, which could therefore be called
DEF. Indeed, the disordering transition resulting in a DEF phase is mostly
promoted by walls, which involve a change of $\pi$ for $\theta$ on either
side of the defect.
On the contrary, $P_{BW}$ is expected to vanish in a hypothetical DOF phase,
since in this case the relevant defects are steps, which involve a
change of $\pm \pi/2$ for $\theta$.\cite{Bernasconi}
The four values of $\theta$ should appear with the same probability in such a
disordered phase, and there is no way of telling which ``color'' prevails
in the top atoms.

The question of possibly finding a DOF phase in the sense of
Ref.\ \cite{Bernasconi} in a model of the clock-step type deserves,
however, a few comments. Suppose that steps were indeed the most energetically
favorable objects in the problem, $E_s\ll E_w$, and imagine desiring
a stable DOF phase, i.e., preventing the appearance of steps from making
the surface immediately rough. The natural way of doing this is to
assign vertex energies to the crossing of steps, in such a way as to
{\em disfavor the crossing of parallel steps\/} with respect to antiparallel
ones.\cite{MdN_Rom} This is indeed the standard mechanism by which a DOF
is stabilized in the context of RSOS models for simple cubic (100) surfaces.
Such 6-vertex energies have been neglected by den Nijs in deriving
the zero-chirality limit phase diagram for the clock-step model.\cite{MdN_92}
It is therefore an appealing suggestion, deserving further study, that their
proper inclusion might open up the possibility of a genuine DOF phase in
the model.\cite{Bernasconi}

Microscopic SOS models are in many ways attractive, at first sight, as
far as stabilizing a DOF is concerned: they automatically tend to disfavor
crossing of parallel steps which involve large height differences.
Moreover, tuning the model parameters offers, in principle, the possibility of
making steps {\em or\/} walls more favorable, at least as far as their $T=0$
energy is concerned.
Things are however not so straightforward in practice.
Consider, as a remarkable counter-example, the case of the $K_4$--model.
When $-1<K_{2x}/K_4<0$, the ground state is $(2\times 1)$ MR
reconstructed, and simple Ising wall defects are energetically more
favorable with respect to all kinds of steps (see table 1).
A value of $K_{2x}/K_4=-0.56$, which we considered in one of the
simulations, would have seemed therefore a quite promising candidate
for a DEF phase.
What we end up with is, instead, a situation quite well represented by the
snapshot in Fig.\ \ref{snapshot:fig}.
The state of the system looks as predicted for a dimer spin state.
We clearly see that there are large regions in which the surface looks
unreconstructed (with either the W or the B sublattice on the top layer),
separated by $(2\times 1)$ steps, forming a fluid with up-down order but
without positional order.
It is worth stressing that the relevant objects in such a disordered phase
-- the $(2\times 1)$ steps -- are the most natural defects of the {\em
unreconstructed\/} surface, and the very building blocks of the neighboring
MR reconstructed surface (which can be seen as a solid of
alternating $(2\times 1)$ steps).
The result of tailoring the $T=0$ defect energies in such a way as to promote
an Ising-wall dominated DEF phase, ends up with an amusing realization of
a dimer state instead.

In conclusion, we believe that a dimer phase type of disordered flat phase
is a natural candidate in systems with a BCSOS-type of symmetry
like the fcc(110) surfaces considered in this work.
Experimental signatures of such a scenario would be the the detection of
a rapid fall of the anti-phase scattering intensity as the critical
temperature is approached, followed by an intermediate temperature region
(before roughening) in which the dominant defects are monoatomic steps, and
where a small coherent anti-phase scattering intensity survives.

ACKNOWLEDGMENTS -- We are grateful to Michele Fabrizio, Marco Bernasconi,
Giorgio Mazzeo, Andrea Levi and Giancarlo Jug for many instructive
discussions.
This research was supported by Italian Research Council (CNR) under the
``Progetto Finalizzato `Sistemi Informatici e Calcolo Parallelo' '' and also
under contract 94.00708.CT02 (SUPALTEMP).
We also acknowledge partial support from EEC contract ERBCHRXCT930342.
G.S. acknowledges support from a ``Human Capital and Mobility''
EEC fellowship during a stay at the European Synchrotron Radiation
Facility (Grenoble, France) where part of the work was done.

%--------------------------------------------------------------------------

%%%%%%%%%%%%%%%%%%%%%%%%%%%%%%%%%%%%%%%%%%%%%%%%%%%%%%%%%%%%%%%%%%%%%%%%%
%                               FIGURES
%%%%%%%%%%%%%%%%%%%%%%%%%%%%%%%%%%%%%%%%%%%%%%%%%%%%%%%%%%%%%%%%%%%%%%%%%
\newpage
\begin{center} {\bf FIGURE CAPTIONS} \end{center}

\begin{figure}
\caption{
Schematic top view of the fcc (110) surface.
The two sublattices, W and B, are denoted by open and solid circles.
In the ideal unreconstructed (110) surface, one of the two sublattices
lies at a distance $a_z=a_y/2$ above the other.
The couplings considered in the model are indicated. Lattice
basis vectors are also shown.
The dashed zig-zag lines represent successive row configurations
(``time-slices'') used in the spin-chain mapping.
}
\label{lattice:fig}
\end{figure}
\begin{figure}
\caption{
Relevant extended defects (steps and walls) of a $(1\times 1)$ and of a
$(2\times 1)$ reconstructed surface.
The ground state energies of these defects are given in table 1.
$\epsilon_{2\times 1}$ is the $(2\times 1)$ (or monoatomic) step discussed
in section \protect\ref{dimer:sec}.
$(2\times 1)$-steps proliferate in the DF phase,
mantaining up-down long-range order.
$\epsilon^*_{2\times 1}$ is a bound pair of $(2\times 1)$-steps, the
relevant defect of an unreconstructed surface.
$\epsilon_{CS}$ and $\epsilon_{AS}$ are clockwise (or $(3\times 1)$) and
anticlockwise (or $(1\times 1)$) steps.
$\epsilon_{Ising}$ and $\epsilon^*_{Ising}$ are two possible types of
domain walls.
}
\label{defects:fig}
\end{figure}
\begin{figure}
\caption{
Schematic height profiles of the two ground states (U1=White, U2=Black) of the
unreconstructed surface, and of the four ground states of the
$(2\times 1)$ missing-row surface.
The reconstruction variable $\theta$ is indicated.
The spin representation of each state, using
Eq.\ \protect\ref{RELAT_H_SPIN:eqn}, is explicitly given.
Notice that the two unreconstructed ground states correspond to the two
possible N\'eel states of a spin-1/2 chain.
}
\label{gs:fig}
\end{figure}
\begin{figure}
\caption{
The exact roughening temperature of the anisotropic BCSOS model
(solid line) and the result obtained by making use of the mapping onto the
XXZ Heisenberg chain (dashed line), as a function of the anisotropy ratio
$K_{2x}/K_{2y}$.
The inset shows the relative discrepancy between the two results.
}
\label{tc_bcsos:fig}
\end{figure}
\begin{figure}
\caption{
Ground state phase diagram of the Heisenberg chain with second
neighbor $S^z_iS^s_{i+2}$ coupling. Ground state degeneracies are given in
square brackets, and the translation of the different phases in the surface
language is explicitly indicated.
The $(1\times 1)$--DF line starting at the point P is continuous, with a
variable exponent.
}
\label{phase_j1j2:fig}
\end{figure}
\begin{figure}
\caption{
Qualitative ground state phase diagram for the Heisenberg chain with second-
and third-neighbor $S^z_iS^s_{j}$ couplings, for $J_3=\alpha J_2$ with
$0< \alpha \le 1/2$. Ground state degeneracies are given in
square brackets. The line labelled `PM' has a variable exponent.
}
\label{phase_j1j2j3:fig}
\end{figure}
\begin{figure}
\caption{
Phase diagram for the $K_3$--model, as obtained from the
quantum spin-chain mapping, for $K_3/K_{2y}=0.025$.
}
\label{phase_k3:fig}
\end{figure}
\begin{figure}
\caption{
Qualitative phase diagram for the modified BCSOS--model in
Eq.\ \protect\ref{model:eqn} for fixed values of the couplings $K_3$ and
$K_4$.
}
\label{phase_k3k4:fig}
\end{figure}
\begin{figure}
\caption{
Finite size behavior of different structure factors at the point
$J_z/J=3.0$ and $J_2/J=2.4$ in the dimer phase: $S^{\rm dim}(\pi)$
(solid squares), $S^{zz}(\pi)$ (crosses), and $S^{zz}(\pi/2)$ (open squares).
The inset shows a logarithmic plot of the dimer structure factor, together
with a dashed line of slope $1$, for comparison.
}
\label{sq:fig}
\end{figure}
\begin{figure}
\caption{
$(2\times 1)$ step-step correlations, see Eq.\ \protect\ref{step_def:eqn},
for a chain of 28 sites, (a) at the point ($J_z=3.0J$,$J_2=2.4J$,$J_3=0$)
inside
the dimer phase, and (b) at ($J_z=3.0J$,$J_2=J_3=0$) inside the N\'eel phase.
For the ideal dimer state $\Psi_1$, see Eq.\ \protect\ref{step_step_corr:eqn},
the correlation function would oscillate between the values $0$ and $1/4$
(the dashed line in (a)).
Inside the N\'eel phase the correlation function decays exponentially to the
square fo the step-density, denoted by the dashed line in (b).
}
\label{step:fig}
\end{figure}
\begin{figure}
\caption{
Finite size scaling of the height fluctuations,
Eq.\ \protect\ref{dh2:eqn}, for the $K_3$ model at
$K_{2x}/K_{2y}=-0.51$, $K_3/K_{2y}=0.22$, $T/K_{2y}=3$.
A line with the critical slope $K(T_R)=1/\pi^2$ is also shown, indicating that
the surface is smooth at this point.
}
\label{auk3_dh2:fig}
\end{figure}
\begin{figure}
\caption{
(a) Finite size scaling of $P_{2\times 1}$ (the reconstruction
order parameter, Eq.\ \protect\ref{P1x1:eqn}, full circles),
$P_{BW}$ (Eq.\ \protect\ref{pbw:eqn}, open diamonds),
and  $P_{BW}^{(2\times 1)}$ (Eq.\ \protect\ref{pbw2x1:eqn}, open squares)
for the $K_3$ model at the same point considered in
Fig.\ \protect\ref{auk3_dh2:fig}.
The inset shows that $P_{2\times 1}$ vanishes as the inverse of the linear
size $L$ of the lattice. The surface is thus deconstructed.
(b) Log-log plot of the finite size behavior of $P_{BW}$, showing the
saturation to a non-zero value for $L\to \infty$.
The surface is in a disordered flat state.
}
\label{auk3:fig}
\end{figure}
\begin{figure}
\caption{
Same as in Fig.\ \protect\ref{auk3:fig}, for a point inside the disordered
flat phase of the $K_4$--model ($K_{2x}/K_{2y}=-0.056$, $K_4/K_{2y}=0.1$,
$T/K{2y}=2.3$).
}
\label{auk4:fig}
\end{figure}
\begin{figure}
\caption{
Snapshot of a surface configuration as generated by the
Monte Carlo simulation for the $K_4$--model at the same point considered in
Fig.\ \protect\ref{auk4:fig}, inside the (dimer) disordered flat phase.
}
\label{snapshot:fig}
\end{figure}
\begin{figure}
\caption{
Sketch of the expected behaviour of the $P_{BW}$ order parameter
(Eq.\ \protect\ref{pbw:eqn}), proportional to the antiphase scattering
intensity, as a funtion of temperature when the preroughening line is crossed.
}
\label{pbw:fig}
\end{figure}

%%%%%%%%%%%%%%%%%%%%%%%%%%%%%%%%%%%%%%%%%%%%%%%%%%%%%%%%%%%%%%%%%%%%%%%%%
%                                TABLE
%%%%%%%%%%%%%%%%%%%%%%%%%%%%%%%%%%%%%%%%%%%%%%%%%%%%%%%%%%%%%%%%%%%%%%%%%
\newpage
\begin{center} {\bf TABLE} \end{center}
\begin{table}
\begin{center}
\begin{tabular}{|l|c|c|}
\hline
                              &$K_3$--model& $K_4$--model   \\ \hline
      $\epsilon_{2\times 1}$  &  $4K_{2x}$ & $4K_{2x}+8K_4$ \\
$\epsilon^{\ast}_{2\times 1}$ &  $8K_{2x}$ & $8K_{2x}+8K_4$ \\ \hline
      $\epsilon_{CS}$     &  $2K_{2x}+8K_3$   &   $2K_{2x}+8K_4$  \\
      $\epsilon_{AS}$     &  $-2K_{2x}$       &   $-2K_{2x}+8K_4$ \\
      $\epsilon_{Ising}$  &  $-4K_{2x}$       &   $-4K_{2x}+8K_4$ \\
$\epsilon^{\ast}_{Ising}$ &  $4K_{2x}+16K_3$  &   $4K_{2x}+16K_4$ \\ \hline
\end{tabular}
\bigskip
\caption[]{Ground state energy of the defects shown
in Fig.\ 2 for the $K_3$-- and $K_4$--model.}
\end{center}
\end{table}


\begin{thebibliography}{10}
\bibitem{Bal_Kar:nota}
{ Balents and Kardar, see Ref.\ \cite{Bal_Kar}, have independently analized the
  fermionic problem obtaining a completely similar phase diagram}.
\bibitem{MdN_Rom:ld}
{ A beautiful discussion of the physics of 1D quantum spin systems and the
  connections to the disordered flat phase of SOS models is given by den Nijs
  and Rommelse in Ref.\ \cite{MdN_Rom}}.
\bibitem{k3_k3k4_diff:nota}
{ The qualitative difference between Figs. \ref{phase_k3:fig} and
  \ref{phase_k3k4:fig} is related to the pathological behavior of the
  $K_3$--model for $K_{2x}\to 0$}.
\bibitem{kin_sign:nota}
{ The sign of the spin-flip term in Eq.\ \ref{SPIN_MOD_J1J2J3:eqn} can be
  changed by the canonical transformation $S^{\pm}_i \to (-1)^i S^{\pm}_i$. A
  singlet pair has the standard sign,
  $(\uparrow\downarrow-\downarrow\uparrow)$, in the new representation, whereas
  it reads $(\uparrow\downarrow+\downarrow\uparrow)$ in the representation in
  which the spin-flip term has a negative matrix element}.
\bibitem{DIFF_MOD:nota}
{ The difference is essentially due to the fact that our second neighbor
  coupling in Eq.\ \ref{SPIN_MOD_J1J2J3:eqn} is of the Ising type
  ($J_2S^z_iS^z_{i+2}$) as opposed to being spin isotropic as it is in the
  $J-J_2$--model. For large values of $J_2$ the model in Eq.\
  \ref{SPIN_MOD:eqn} with $J_3=0$ maps onto two uncoupled antiferromagnetic
  Ising chains. Given the large gap of the Ising problem, a small nearest
  neighbor XXZ term has no effect on it and the systems remains ordered. On the
  contrary, the spin-isotropic model in Eq.\ \ref{HAL_MOD:eqn} maps, for large
  $J_2$, onto two uncoupled isotropic Heisenberg chains, which are gapless and
  for which the nearest neighbor XXZ term is a relevant perturbation}.
\bibitem{step_3x1:nota}
{ It is worth mentioning here what happens to $(3\times 1)$ steps, i.e., three
  neighboring like-spins ($\uparrow\uparrow\uparrow$ for up-steps and
  $\downarrow\downarrow\downarrow$ for down-steps). Such objects -- strictly
  absent in the two representative dimer states in Eq.\ \ref{DIMERS:eqn} -- are
  quite rare in the dimer phase and play no role: their correlation function
  decays exponentially to the square of their concentration without any form of
  LRO}.
\bibitem{PBW_PEO:nota}
{ $P_{BW}$ coincides, apart from a factor $2$, with the order parameter
  $P_{EO}$ defined in Ref.\ \cite{Bernasconi}. To avoid possible
  misunderstandings, we mention that in Ref.\ \cite{Mazzeo} $P_{BW}$ denotes
  the order parameter of the unreconstructed surface $P_{(1\times 1)}$}.
\end{thebibliography}
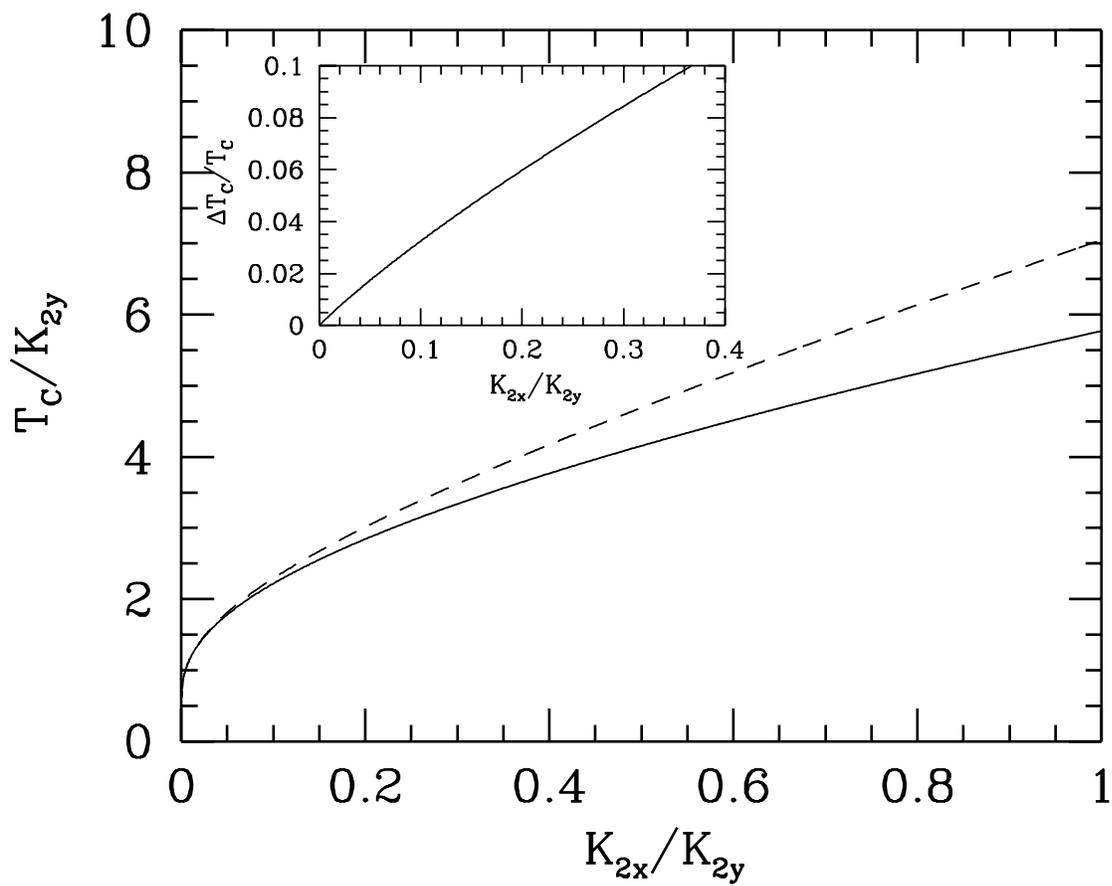
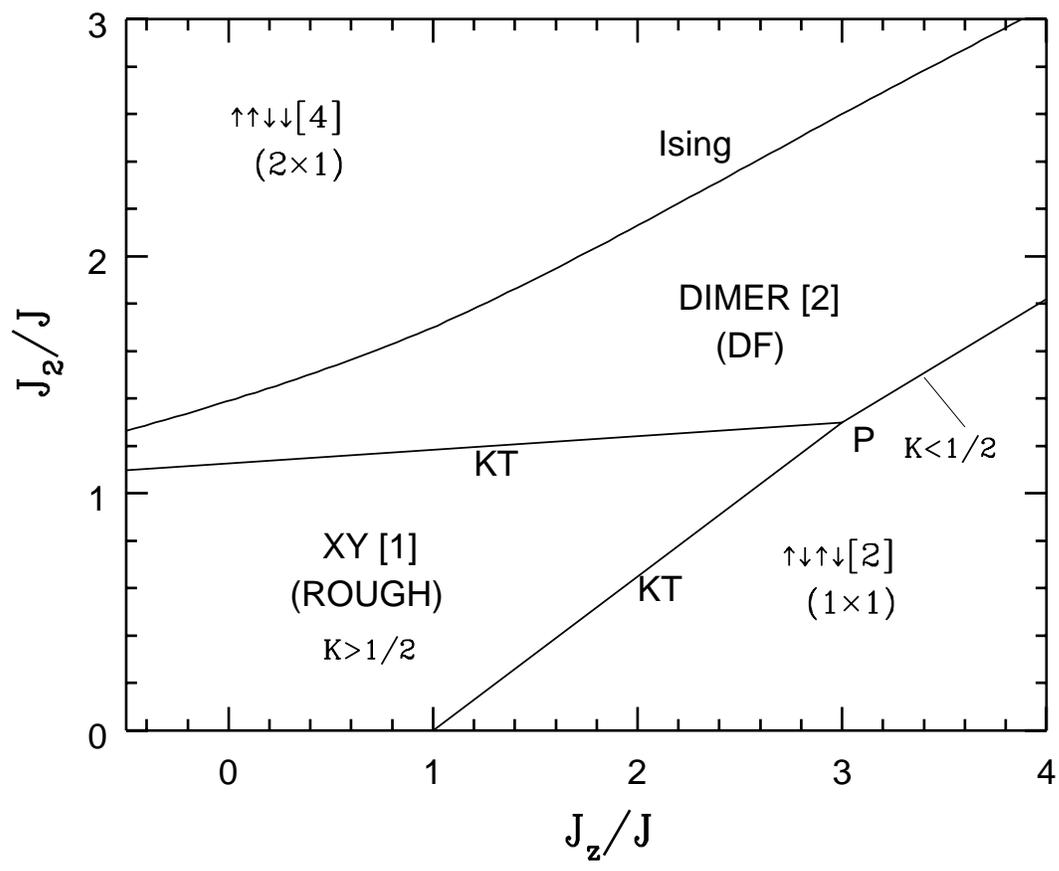
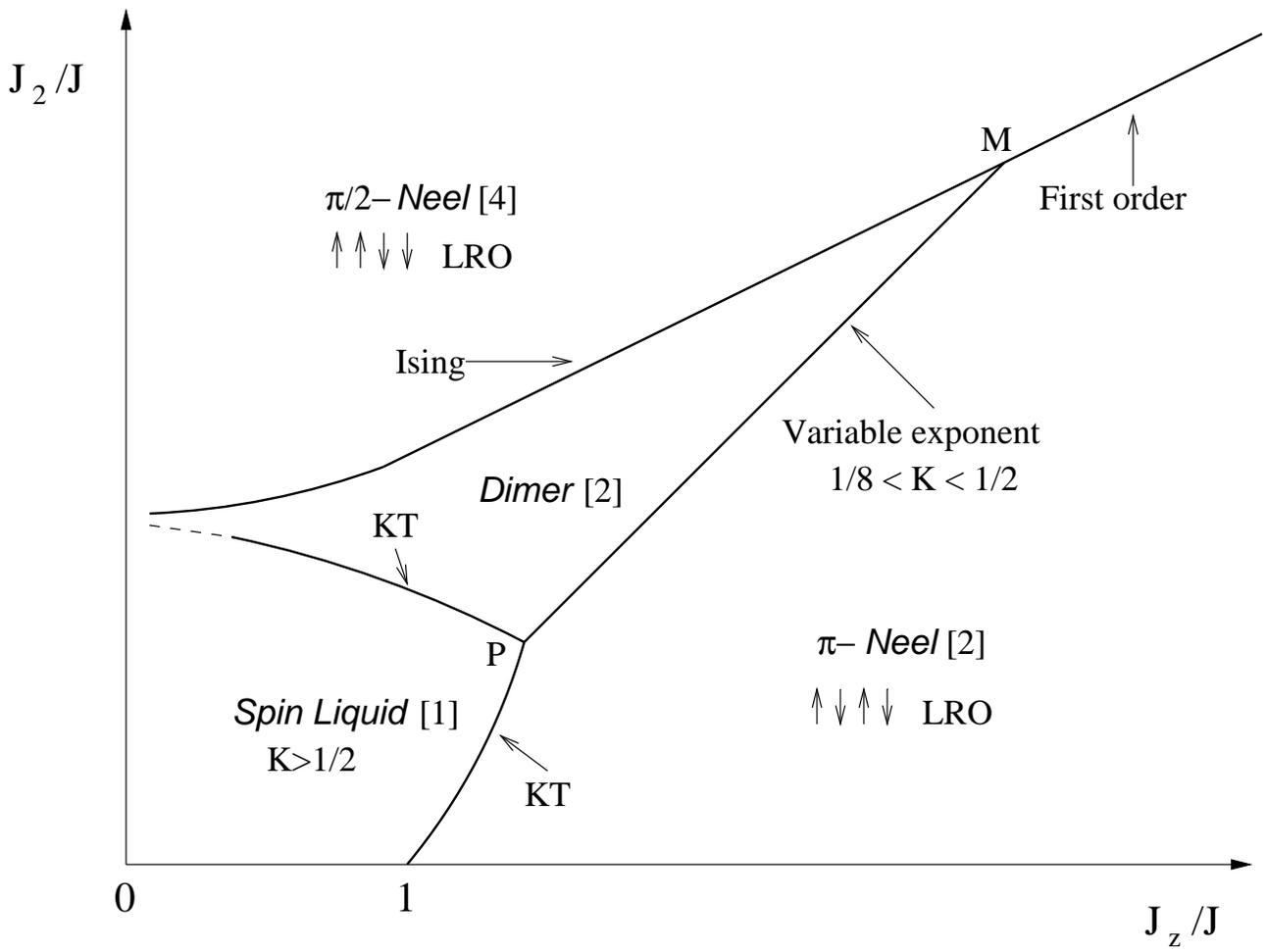
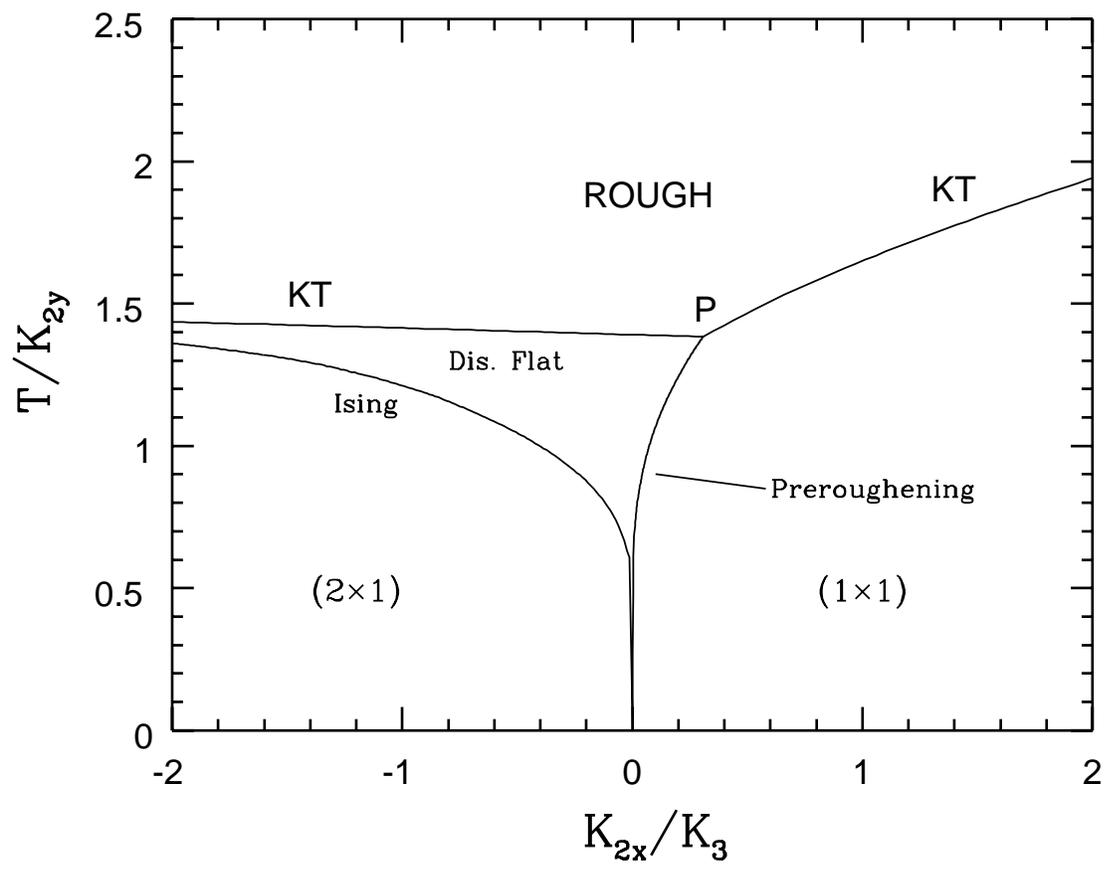
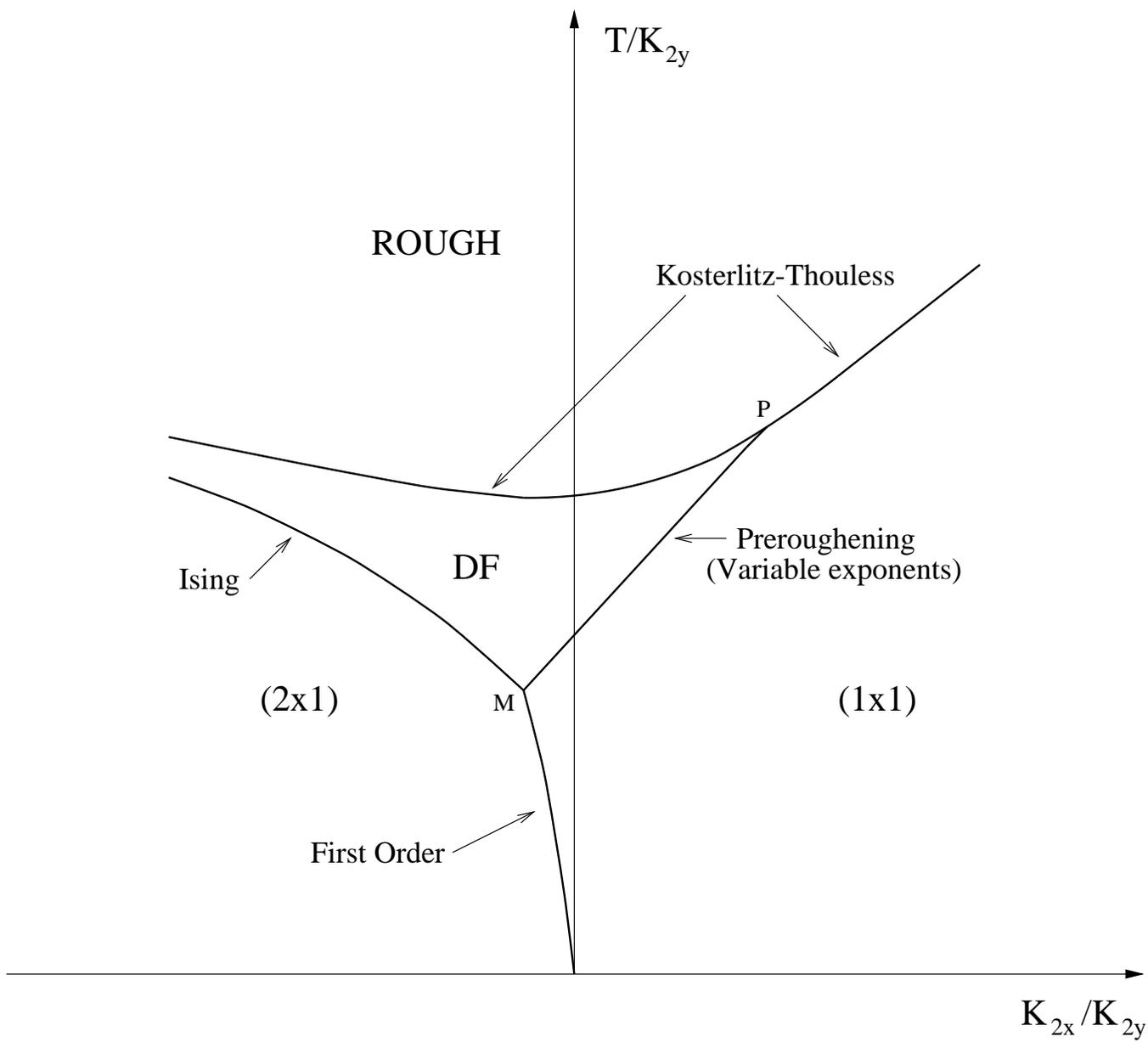
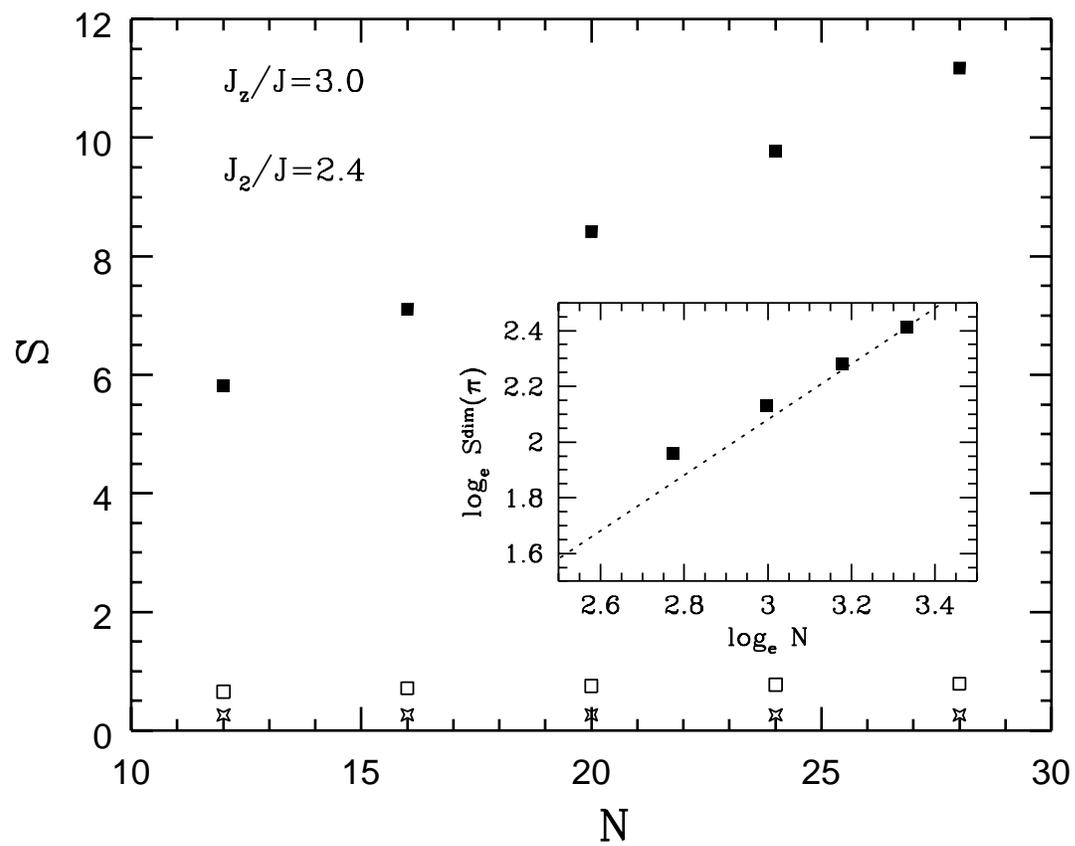
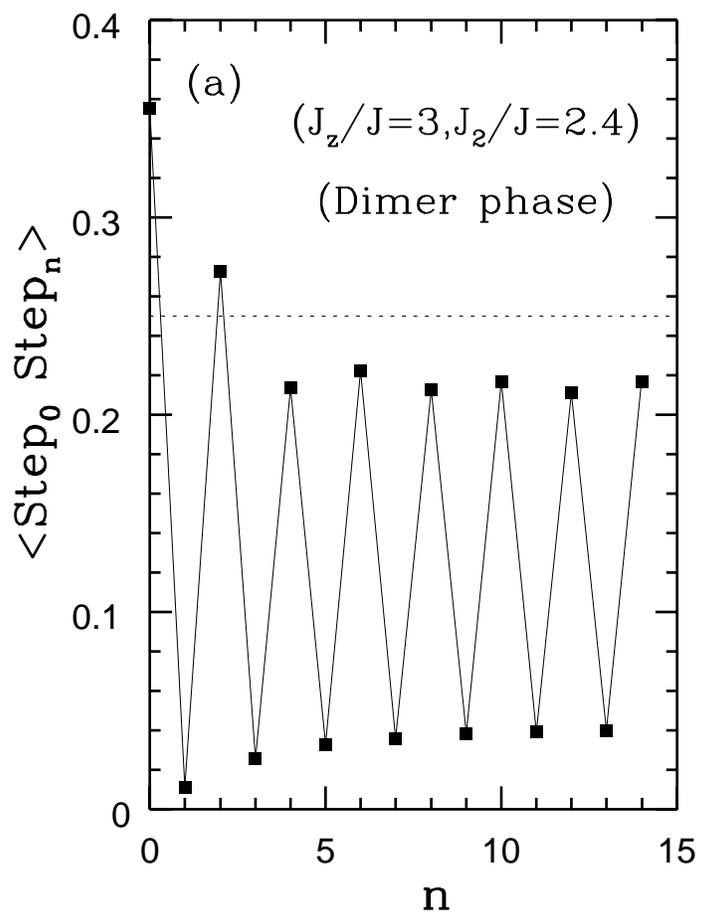
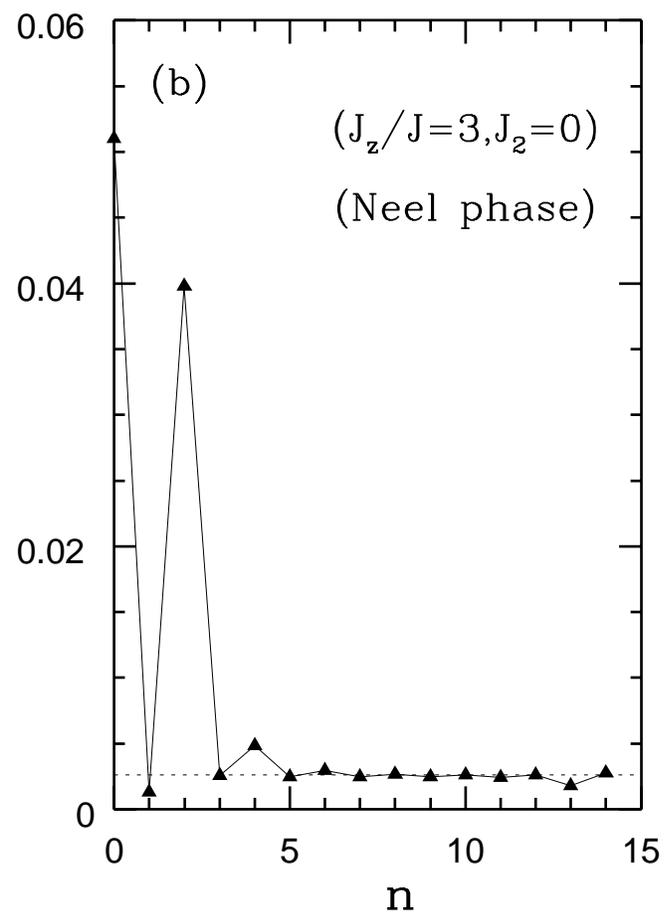
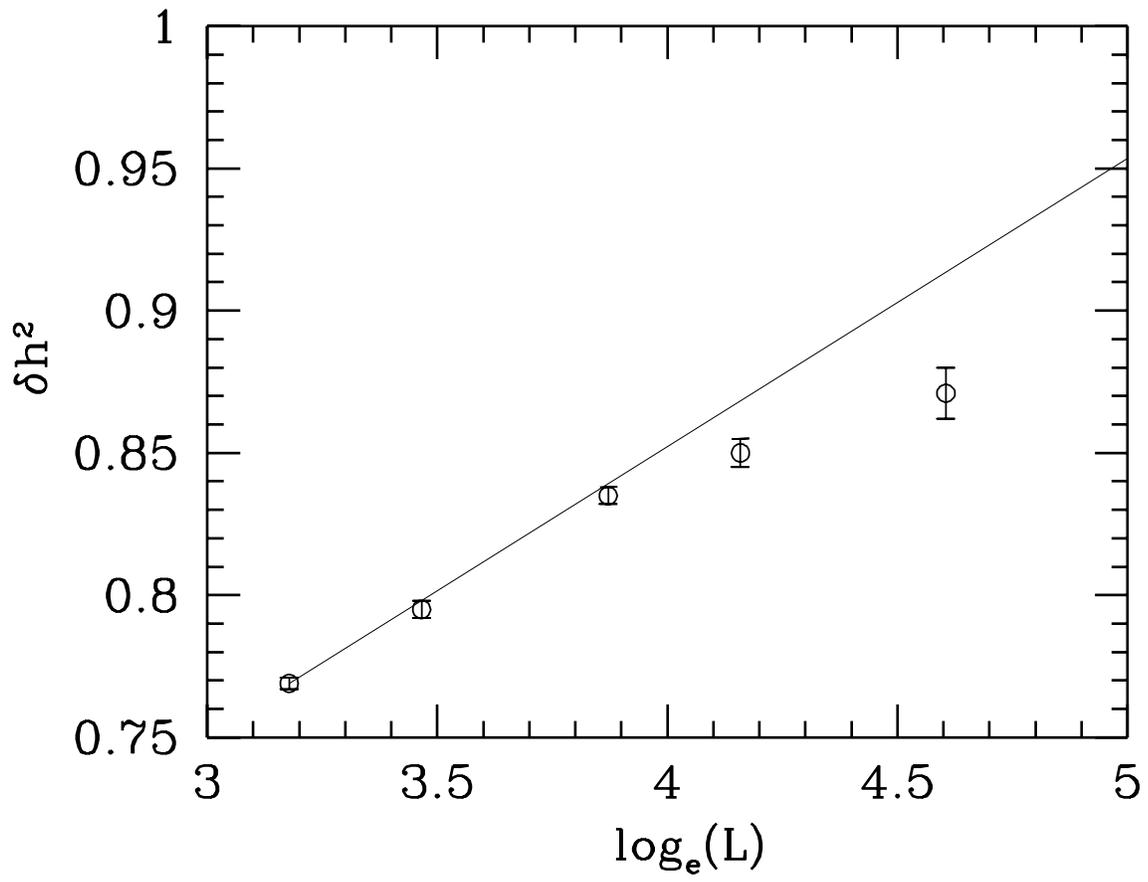
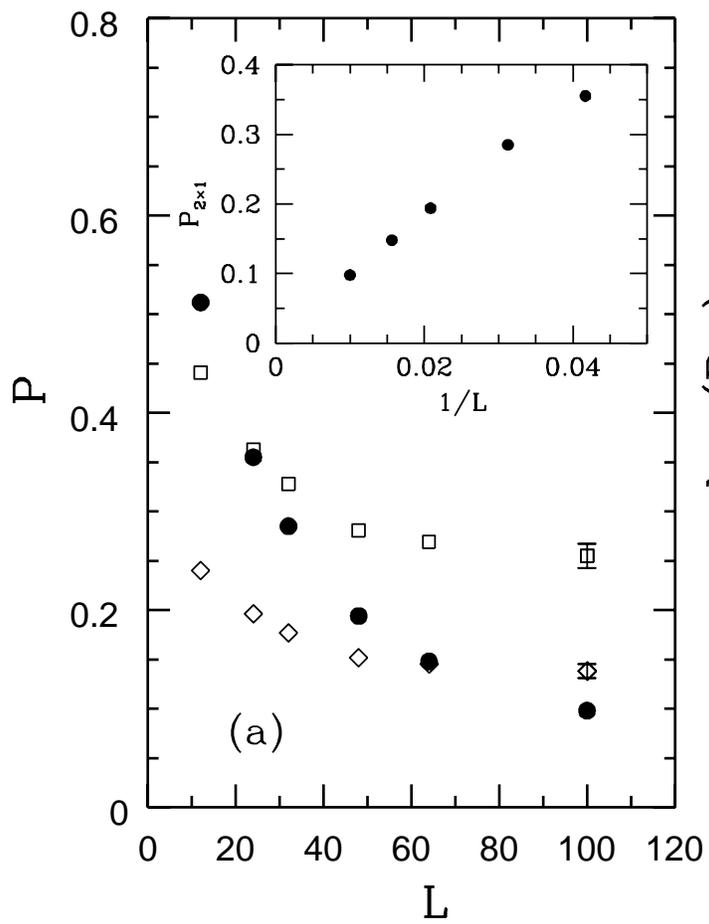
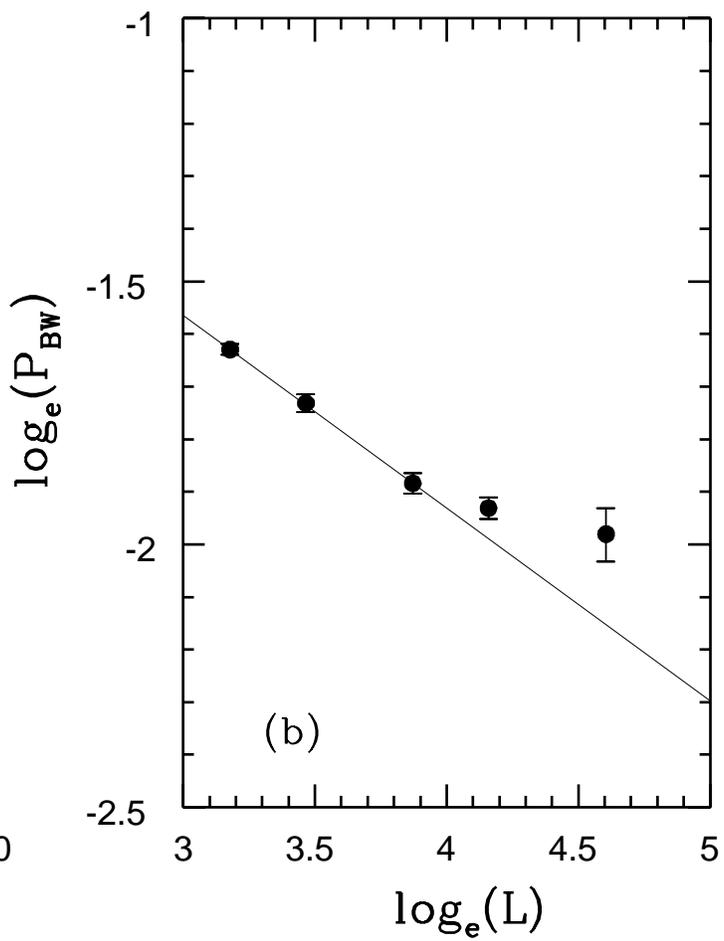
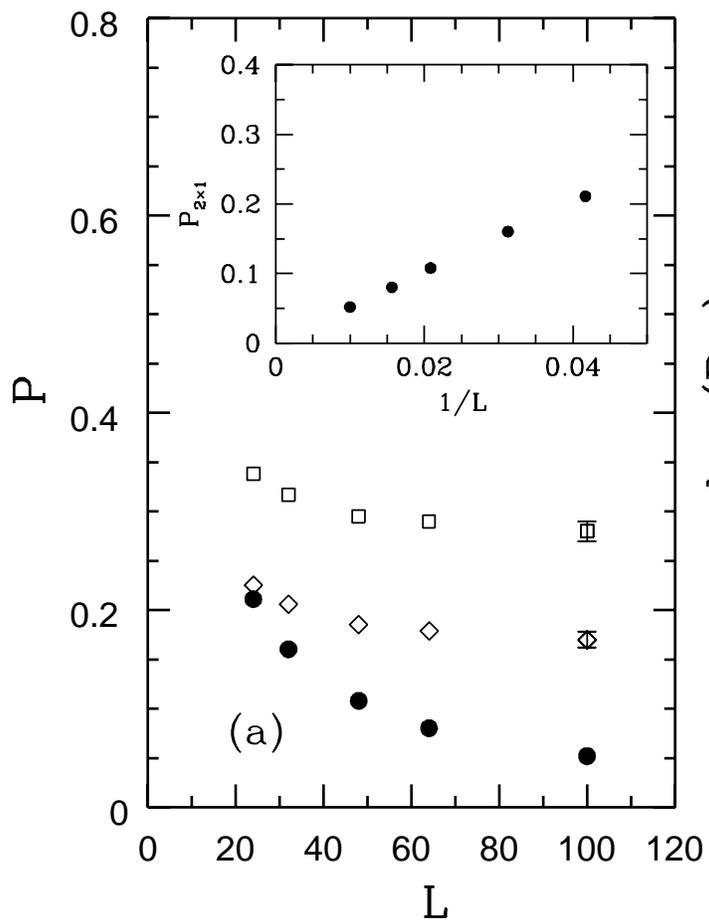
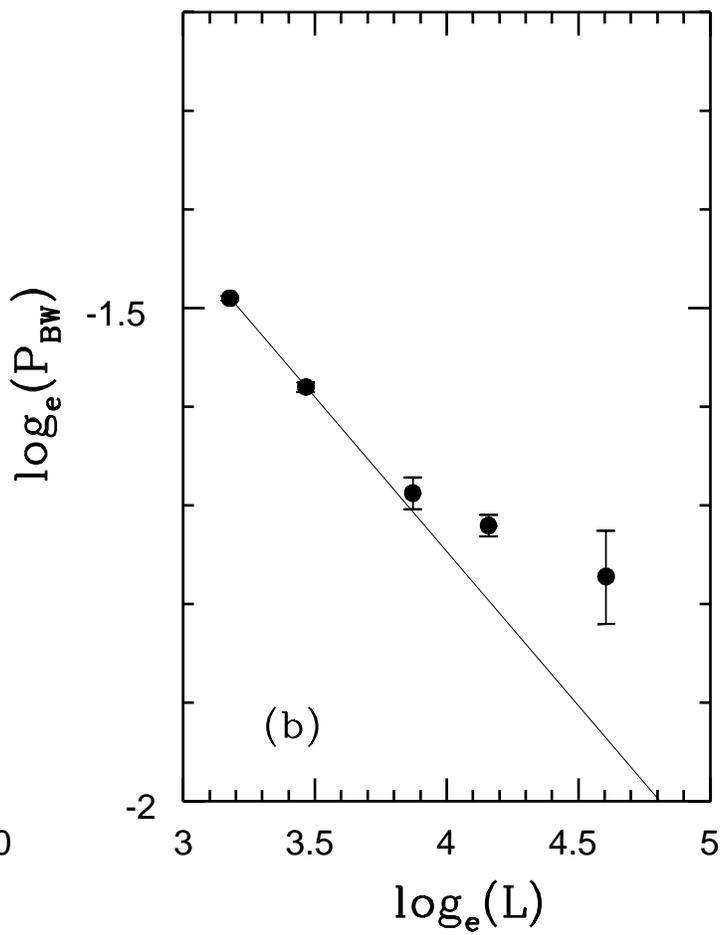
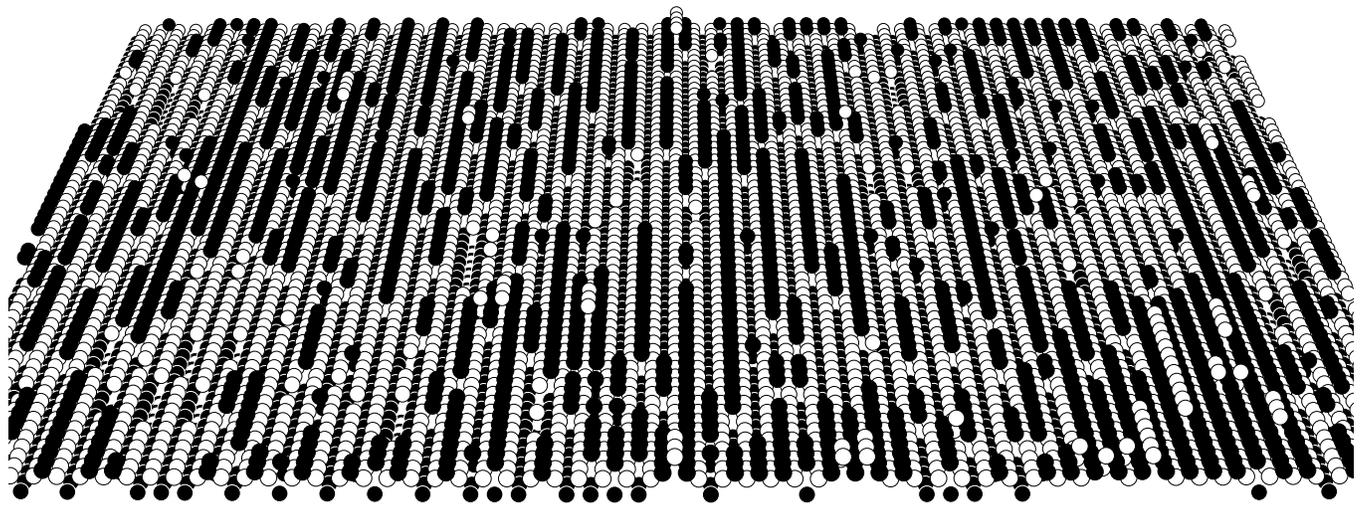
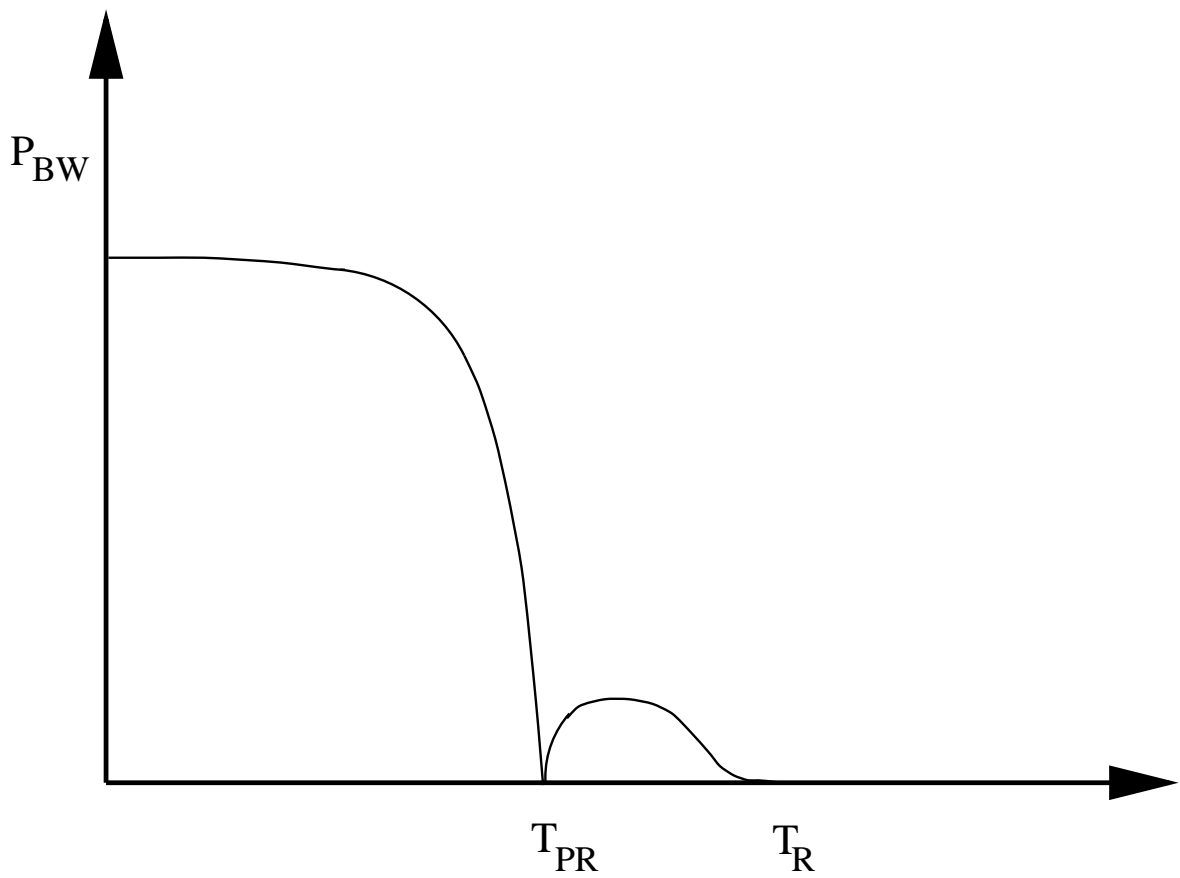
\end{document}